\documentclass[12pt,preprint]{aastex}
\usepackage{emulateapj5}
\usepackage{apjfonts}
\usepackage{amssymb}
\usepackage{natbib}
\usepackage{lscape}

\newenvironment{inlinefigure}{%
\def\@captype{figure}%
\vspace{0.15in}%
\noindent\begin{minipage}{0.999\linewidth}\begin{center}\small}
{\end{center}\end{minipage}\smallskip}
\makeatother

\newcommand{\msun}{\,{\rm M_\odot}}
\newcommand{\Ha}{\,{\rm H\alpha}}
\newcommand{\Hb}{\,{\rm H\beta}}
\newcommand{\kms}{km~s$^{-1}\,$}

\begin{document}

\title{H$\mathbf{\alpha}$ SPECTROSCOPY OF GALAXIES AT \lowercase{$z>2$}:
  KINEMATICS AND STAR FORMATION\altaffilmark{1}}
\author{\sc Dawn K. Erb, Alice E. Shapley, Charles C. Steidel\altaffilmark{2}}
\affil{California Institute of Technology, Department of Astronomy, MS
  105-24, Pasadena, CA 91125} 
\author{\sc Max Pettini}
\affil{Institute of Astronomy, Madingley Road, Cambridge CB3 0HA, UK}
\author{\sc Kurt L. Adelberger\altaffilmark{3}}
\affil{Harvard-Smithsonian Center for Astrophysics, 60 Garden Street,
  Cambridge, MA 02138} 
\author{\sc Matthew P. Hunt}
\affil{California Institute of Technology, Department of Astronomy, MS
105-24, Pasadena, CA 91125}
\author{\sc Alan F.M. Moorwood}
\affil{European Southern Observatory, Karl-Schwarzschild-Str. 2, 
D-85748 Garching, Germany}
\and
\author{\sc Jean-Gabriel Cuby}
\affil{European Southern Observatory, Alonso de Cordova 3107,
Santiago, Chile}
\email{dke@astro.caltech.edu}

\submitted{Accepted for publication in \apj}

\altaffiltext{1}{Based, in part, on data obtained at the 
W.M. Keck Observatory, which is operated as a scientific partnership
among the California Institute of Technology, the University of
California, and NASA, and was made possible by the generous financial
support of the W.M. Keck Foundation.  Also based in part on observations 
collected at the European Southern Observatory, Paranal, Chile (ESO 
Programme 66.A-0206).}  
\altaffiltext{2}{Packard Fellow}
\altaffiltext{3}{Harvard Society Junior Fellow}

\begin{abstract}
We present near-infrared spectroscopy of $\Ha$ emission lines in a sample of
16 star-forming galaxies at redshifts $2.0 < z < 2.6$.  
Our targets are drawn from a large sample of galaxies photometrically
selected and spectroscopically confirmed to lie in this redshift
range. We have obtained this large sample with an extension of the broadband
$U_nG\cal R$ color criteria used to identify Lyman break galaxies at
$z\sim3$.  The primary selection criterion for IR spectroscopic observation was
proximity to a QSO sight-line; we therefore expect the galaxies
presented here to be representative of the sample as a whole.  Six of
the galaxies exhibit spatially extended, tilted $\Ha$ emission lines; rotation
curves for these objects reach mean velocities of $\sim150$ \kms at
radii of $\sim6$ kpc, without corrections for inclination or any other
observational effect.  The velocities and radii give a mean dynamical
mass of $\langle M \rangle \geq 4 \times 10^{10} \msun$.  We have
obtained archival \textit{HST} images for two of these galaxies; they
are morphologically irregular.  One-dimensional velocity dispersions
for the 16 galaxies range from $\sim50$ to $\sim260$ \kms, and in
cases where we have both virial masses implied by the velocity
dispersions and dynamical masses derived from the spatially extended
emission lines, they are in rough agreement.  We compare our kinematic
results to similar measurements made at $z\sim3$, and find that both
the observed rotational velocities and velocity dispersions tend to be larger
at $z\sim2$ than at $z\sim3$. 
We also calculate star
formation rates (SFRs) from the $\Ha$ luminosities, and compare them
with SFRs calculated from the UV continuum luminosity. We find a mean
$\rm SFR_{\Ha}$ of 16 $\msun \; \rm yr^{-1}$ and an average $\rm
SFR_{\Ha}/SFR_{UV}$ ratio of 2.4, without correcting for extinction.
We see moderate evidence for an inverse correlation between the UV
continuum luminosity and the ratio $\rm SFR_{\Ha}/SFR_{UV}$, such as
might be observed if the UV-faint galaxies suffered greater
extinction.  We discuss the effects of dust and star formation history
on the SFRs, and conclude that extinction is the most likely
explanation for the discrepancy between the two SFRs.
\end{abstract}


\section{Introduction}

Our knowledge of star-forming galaxies at high redshift has increased
enormously in the past ten years, particularly at $z\sim3$; large samples
of galaxies at these redshifts are now known \citep{sag+99,sas+03},
and they have been studied in both the rest-frame UV
\citep{psa+00,ssp+03} and optical (\citealt{ssa+01}; \citealt*{pdf01};
\citealt{pss+01}), as well
as at submillimeter \citep{css+00,as00} and X-ray
\citep{nma+02} wavelengths to some extent.  Much less is known about
galaxies at
$z\sim2$.  Because these objects lack strong spectroscopic features in
the optical window, they have traditionally been difficult to
identify.  This is unfortunate, as $z\sim2$ is likely the epoch in which a
large fraction of the stars in the present day universe formed
(\citealt*{mpd98}; \citealt{bsik99}), in which bright QSO activity
reached its peak (\citealt*{ssg95,p95}; \citealt{fss+01}), and in which rapidly
star-forming galaxies of compact and disordered morphologies became
the normal Hubble sequence galaxies of the $z<1$ universe \citep{d00}.

The situation is improving, however.  With the advent of sensitive IR
detectors observations of rest-frame optical features are now feasible, and
have been carried out successfully.  \citet*{tmm98} reported 11 $\Ha$
emitters discovered in a narrow-band IR imaging survey; \citet{ymf+99}
and \citet*{hcs00} used slitless spectroscopy with the Near Infrared
Camera and Multi-Object Spectrograph (NICMOS) on the \textit{Hubble
  Space Telescope (HST)} to study the $\Ha$ luminosity function and
star formation rate in galaxies at $z\leq1.9$.  Objects at
$z\sim2$ are in fact ideally suited for ground-based IR spectroscopy, since
$\Ha$ falls in the $K$-band, [\ion{O}{3}] and $\Hb$ in the $H$-band,
and [\ion{O}{2}] in the $J$-band.  This coincidence has been exploited with
recent observations employing near-IR spectrographs on 8--10 m
telescopes; most of these have focused on $\Ha$ emission \citep{kk00,lcp+02}. 
Among these spectra is a rotation curve of a galaxy at $z\sim2$
that reaches a velocity of $\gtrsim200$ \kms \space \citep{lcp+02},
suggesting that near-IR spectroscopy may be able to provide the most
detailed kinematic information yet available on galaxies at high
redshift.  It is also clear from most of the above results
that star formation rates measured from $\Ha$ are consistently higher
than those measured from the UV continuum luminosity; this is
in accordance with observations at $z\sim1$ \citep{gbe+99,tmlc02}
and at lower redshifts (e.g.\ \citealt{bk01}; also see \citet{ste+00}
and \citet{bbgb02} for comparisons of $\Ha$ and UV SFRs).  The
difference is generally accounted for by the differing sensitivities
of the $\Ha$ and UV continuum star formation rate diagnostics to the
presence of dust and to star formation history.

In this paper we present $\Ha$ spectroscopy in the $K$-band of 16
UV-selected galaxies in the redshift range $2.0 < z < 2.6$.  In
\S~\ref{sec:obs} we describe our target selection process,
observations, and data reductions.  In \S~\ref{sec:ind} we comment
individually on any noteworthy features of the galaxies.  \S~\ref{sec:kin}
addresses the kinematics of the galaxies:  we discuss the rotation
curves in \S~\ref{sec:rot} and the one-dimensional velocity
dispersions in \S~\ref{sec:veld}.  In \S~\ref{sec:sfrs} we calculate
star formation rates from $\Ha$ and rest-frame UV emission and compare
them, and we discuss our conclusions in \S~\ref{sec:disc}.  We use a
cosmology with $H_0=70\;{\rm km}\;{\rm s}^{-1}\;{\rm Mpc}^{-1}$,
$\Omega_m=0.3$, and $\Omega_{\Lambda}=0.7$ throughout.  In this
cosmology, the universe at $z=2.3$ is 2.8 Gyr old, or 21\% of its
present age, and a proper distance of 8.2 kpc subtends an angular
distance of 1\arcsec. 

\section{Target Selection and Observations}
\label{sec:obs}

The objects discussed herein are drawn from a large sample of galaxies
photometrically selected and spectroscopically confirmed to be in the
redshift range $2.0 \le z \le 2.6$.  We summarize the
selection technique here; a more complete discussion will be given in
a forthcoming paper.  We have extended the
broadband color criteria used to 
select galaxies at $z\sim3$ (\citealt{sh93}; \citealt*{sph95};
\citealt{sgp+96})  to other regions
of the $(U_n-G)$ vs. $(G-\cal R)$ plane, identifying candidates
according to the following conditions:
\begin{eqnarray}
G-{\cal R} &\geq& -0.1\nonumber\\
U_n-G      &\geq& G-{\cal R}+0.2\\
G-{\cal R} &\leq &0.3(U_n-G)+0.2\nonumber\\
U_n-G      &<& G-{\cal R}+1.0\nonumber
\end{eqnarray}
We refer to these objects as ``BX'' (e.g.\ Q1700-BX691); 92\%
of the objects satisfying these criteria are galaxies in the redshift
range $1.6 \le z \le 2.8$, with 72\% in the range $2.0 \le z \le 2.6$.
These criteria were developed by calculating the colors that typical
$z\sim3$ Lyman break galaxies (LBGs) would have if they were placed at
$z\sim2$; they are therefore designed to select objects with similar
intrinsic spectral energy distributions (SEDs) at both redshifts
\citep{a02}.  Our sample
also contains four ``MD'' objects (e.g.\ Q1623-MD107); these objects are
detected in the $U_n$-band and meet the criteria
\begin{eqnarray}
G-{\cal R} &<& 1.2\nonumber\\
U_n-G &\leq& G-{\cal R}+1.5\\
U_n-G &\geq& G-{\cal R}+1.0\nonumber
\end{eqnarray}
They have the redshift distribution $\langle z \rangle = 2.79\pm0.27$
\citep{sas+03}, so that the low redshift end of the distribution
encompasses objects with $z \leq 2.6$.  Both the BX and MD candidates are
restricted to ${\cal R}\leq25.5$ (roughly equivalent to ${\cal
  R}\lesssim26$ at $z\sim3$).  The two remaining objects
in our sample, Q0201-B13 and CDFb-BN88, satisfy the BX criteria but
have different names because they predated the systematic use of the
$z\sim2$ selection technique.  Once candidates are photometrically
identified, we confirm their redshifts with rest-frame UV spectra
obtained with the Low Resolution Imaging Spectrometer (LRIS;
\citealt{occ+95}) on the Keck I telescope.  The redshifts from the UV
interstellar absorption lines and Ly$\alpha$ when present are listed
in Table~\ref{tab:obs}, and 
spectra for two of the objects are shown as examples in
Figure~\ref{fig:uvspec}.  The rest-frame UV observations will be
described in detail elsewhere. 

\begin{inlinefigure}
\centerline{\epsfxsize=9cm\epsffile{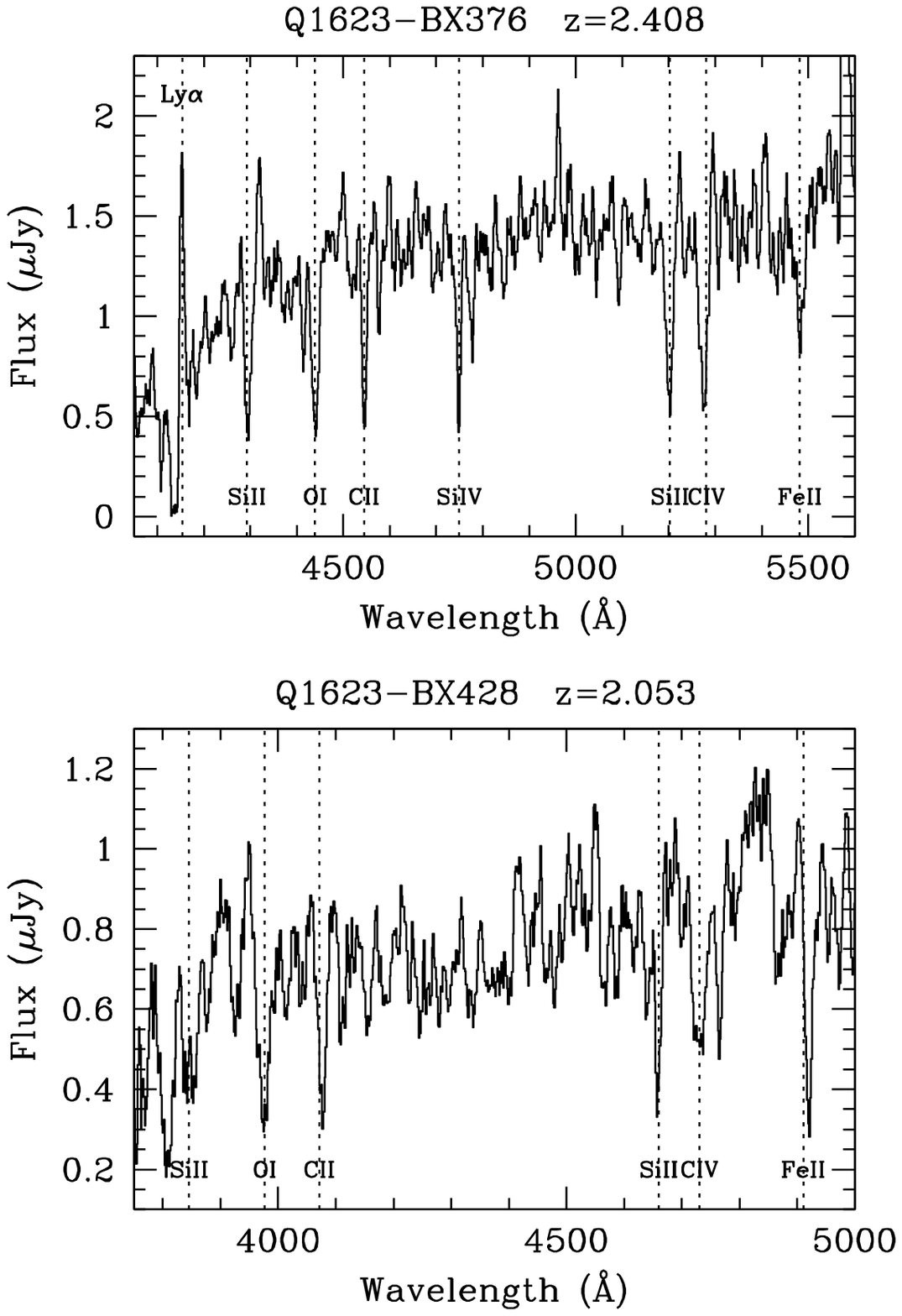}}
\figcaption{Sample rest frame UV spectra for two of the galaxies.  We
  show Q1623-BX376 
  at $z_{\rm abs}=2.408$ (top), and Q1623-BX428 at $z_{\rm abs}=2.053$
  (bottom).  The rest wavelengths of the lines labeled are
  Ly$\alpha$ $\lambda$1215 \AA, SiII $\lambda$1260 \AA, OI $\lambda$1302
  \AA, CII $\lambda$1334 \AA, SiIV $\lambda$1394 \AA, SiII $\lambda$1526 \AA,
  CIV $\lambda$1549 \AA, and FeII $\lambda$1608 \AA. 
\label{fig:uvspec}
}
\end{inlinefigure}


\begin{figure*}
\centerline{\epsfxsize=18cm\epsffile{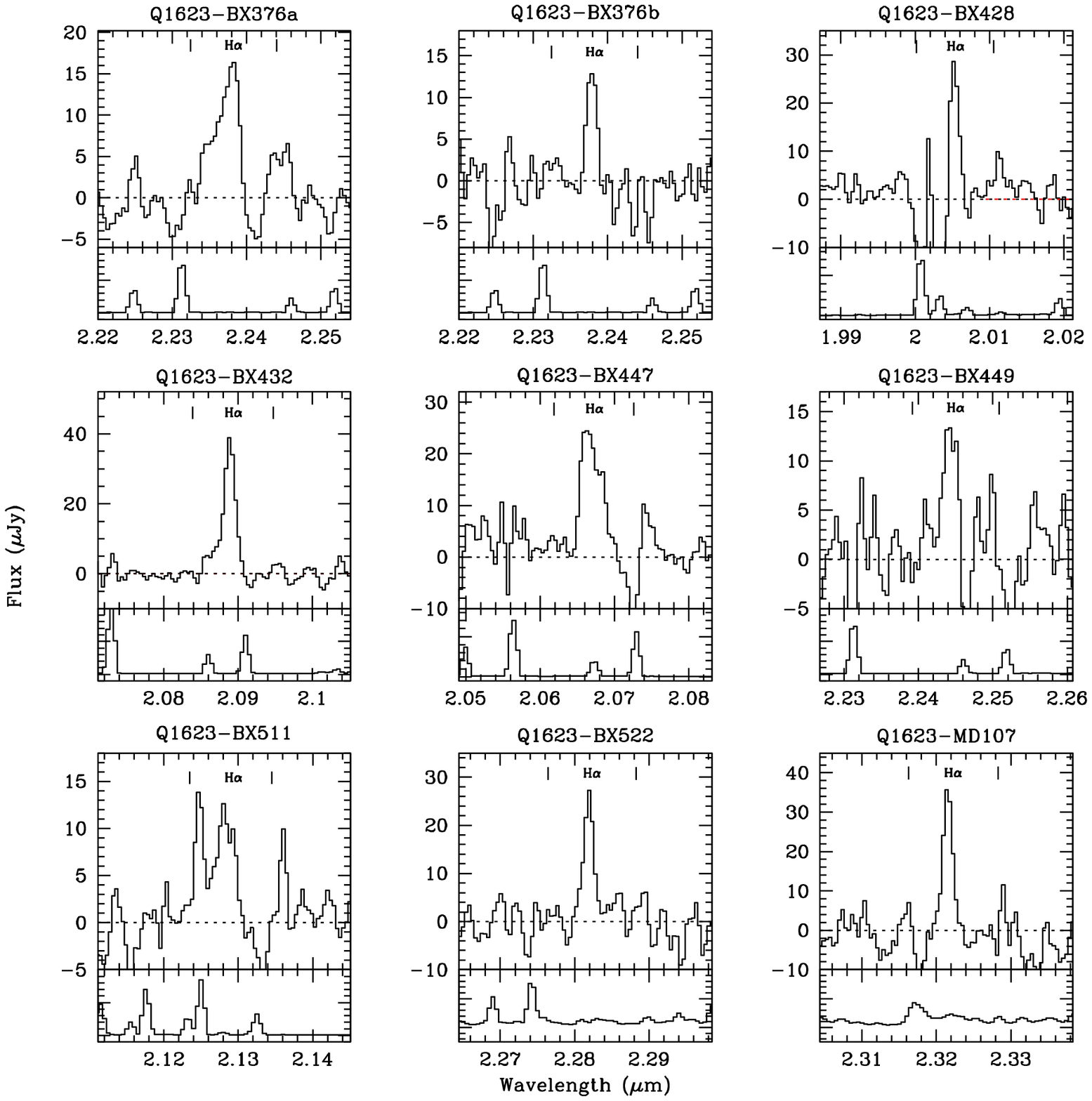}}
\figcaption{Fully reduced one-dimensional spectra for all of the galaxies
in our sample.  The $\Ha$ emission line is marked on each spectrum,
and the vertical lines to either side mark the positions at which
[\ion{N}{2}] emission would appear.  Plotted below each galaxy
spectrum is a sky spectrum, in arbitrary flux units. The spectra have
been smoothed with a two pixel boxcar filter.  We discuss the objects
individually in \S~\ref{sec:ind}.
\label{fig:specs}
}
\end{figure*}

\begin{figure*}
\addtocounter{figure}{-1}
\centerline{\epsfxsize=18cm\epsffile{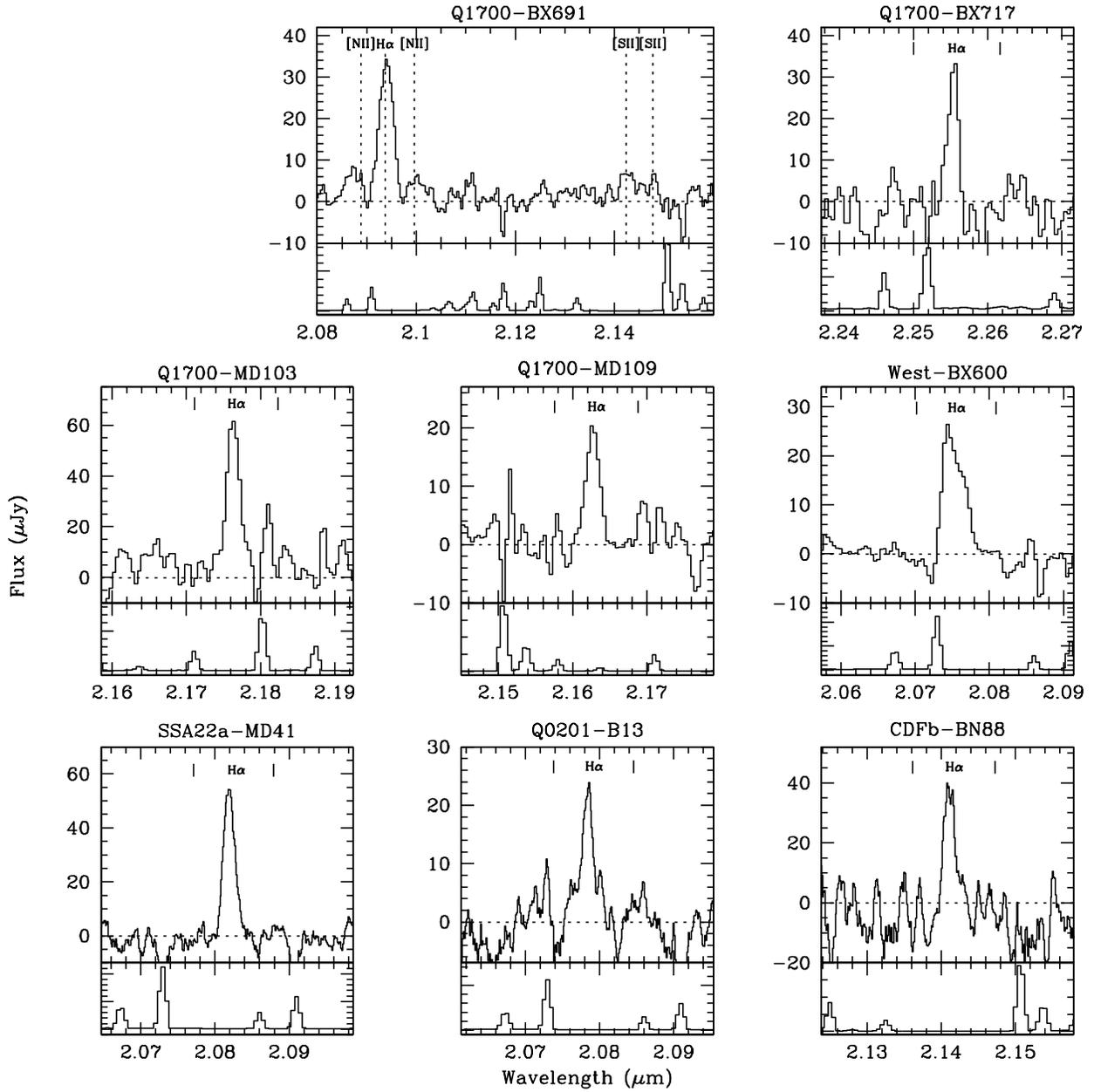}}
\figcaption{continued. We plot a larger
wavelength range for Q1700-BX691, the only object in which we see
[\ion{N}{2}] and [\ion{S}{2}] emission.  The last three objects, SSA22a-MD41,
  Q0201-B13, and CDFb-BN88, were observed with ISAAC on the VLT, and
  their spectra have been smoothed to approximate the resolution of
  NIRSPEC.}
\end{figure*}

The galaxies targeted for IR spectroscopy were selected as part of an
ongoing project examining the interplay between galaxies and the
intergalactic medium (IGM) in which we combine spectroscopy of faint
star-forming galaxies with QSO absorption line observations of the IGM in
the same volume \citep{assp03}.  A detailed comparison of the galaxies and
the IGM requires accurate measurements of the galaxy redshifts, and
ultimately an understanding of the star formation rates, masses and
ages of galaxies near the QSO lines of sight; therefore the primary
selection criterion (beyond the color criteria described above) for
the present sample was proximity to a QSO sight-line.  This naturally
results in 
a sample with a wide range of UV properties (as distinguished, for
example, from the galaxies in the $z\sim3$ sample of \citet{pss+01},
which were selected to be particularly UV-bright). 

Twelve of our sixteen galaxies are within 60\arcsec \space of QSOs in fields
at 1700+64 and 1623+27, and have redshifts slightly lower than those of
the QSOs themselves; these were observed with the Near Infrared
Imaging Spectrograph (NIRSPEC; \citealt{mbb+98}) on the Keck II
telescope in May 2002.  We observed an additional galaxy in the
Groth-Westphal field on the same run.  The other three objects in the sample
(SSA22a-MD41, Q0201-B13, and CDFb-BN88) were observed with the
Infrared Spectrometer and Array Camera (ISAAC; \citealt{mcb+98}) on the
Very Large Telescope 1 (VLT 1) in October 2000, and were among
the small number of $z\sim2$--2.5 galaxies in the $z\sim3$ LBG survey
fields at the time.  They were also selected because 
of their UV brightness, and because of the favorable wavelength of
$\Ha$ relative to night sky emission lines and the possibility of
measuring rotation.

\subsection{Data Acquisition}
\label{sec:dacq}

Most of our targets were observed on May 19 and 20, 2002
(UT) with the NIRSPEC spectrograph on the Keck II telescope.  NIRSPEC is
described in detail by \citet{mbb+98}; it uses a 1024 $\times$
1024 pixel (ALADDIN2) InSb detector with 27 \micron \space pixels.  In the
medium-dispersion mode used for these observations, each detector
pixel corresponds to 0\farcs 143 in the spatial direction, and the
dispersion in the spectral direction is 4.2 \AA \space per pixel.  We used
a 0\farcs 76 $\times$ 42\arcsec \space entrance slit, which gives a
resolving power of $R \simeq 1400$ corresponding to a spectral 
resolution of $\sim 15$ \AA \space FWHM in the observed frame
$K$-band, as measured
from the widths of sky lines.  In almost all cases we were able to place
two galaxies on the slit at the same time by setting the appropriate
position angle.  Because the galaxies are too faint to be acquired
directly on the spectrograph slit, we placed them on the slit by
offsetting from a nearby bright star or from the QSO with a sight-line
near the galaxy.  Individual
exposures were 900 s, and we typically took four exposures of each
object for a total of 1 hour of integration.  Between each exposure we
reacquired the offset star, moved it along the slit by approximately
5\arcsec, and offset once again to the target object.  The detector
was read out in multiple-read mode, with 16 reads at the start and end
of each integration; the results were then averaged to reduce noise.
The choice of filter and wavelength range was governed by the expected
position of the $\Ha$ line based on each galaxy's optical redshift; we
used the NIRSPEC6 and NIRSPEC7 filters, which span the wavelength
ranges 1.56--2.32 and 1.84--2.63 \micron \space respectively.  The
spectral  dispersion allows a range of approximately 0.4 \micron
\space to be placed on the detector at one time.  Conditions were
photometric on both nights, with approximately 0\farcs 5 FWHM seeing
in $K$-band. 

SSA22a-MD41, Q0201-B13, and CDFb-BN88 were observed on October 20--22,
2000 (UT) with the ISAAC spectrograph on the VLT1.  The
short-wavelength channel of ISAAC \citep{mcb+98} uses a 1024
$\times$ 1024 pixel Rockwell HgCdTe array with 18.5 \micron \space pixels.
The pixel scale along the 1\arcsec $\times$ 120\arcsec \space slit is similar
to that of NIRSPEC, 0\farcs 146 \space pixel$^{-1}$, but the spectral
resolution is 2.5 times higher, with $R \simeq 3500$ and sky line
widths of $\sim6$ \AA
\space FWHM.  We observed in the $K$-band, again targeting the
expected position of $\Ha$ from rest-frame UV redshifts.  The position
angles were chosen to align with the major axes of the galaxies if any
extended structure was apparent in the optical images; this was the
case with SSA22a-MD41 and with CDFb-BN88 to a lesser extent.  We also
placed a bright star on the slit along with each galaxy to facilitate
the determination of offsets between images.  We performed an ABBA
series of four 720 second exposures, with 10\arcsec \space offsets
between the A and B positions.  The object was then reacquired at a
different position along the slit and the procedure was repeated,
typically for a total of $\sim3$ hours of integration.  Conditions
were not photometric, and the seeing varied between 0\farcs 5 and
0\farcs 6 FWHM.  The targets and observations are summarized in
Table~\ref{tab:obs}. 

\subsection{Data Reduction}
\label{sec:redux}

The fully reduced spectra are shown in Figure~\ref{fig:specs}.
The two-dimensional images were reduced with IRAF; preliminary steps
included flagging and masking any pixels that exhibited aberrant
behavior in the dark and flat-field images,
flat-fielding the data using the spectrum of a quartz
halogen lamp, and cutting out and rotating the image of the slit.
Spatial distortion was corrected by stepping a bright star
along the slit at 5\arcsec \space intervals for the NIRSPEC data and
10\arcsec \space intervals for that from ISAAC, combining the resulting
images, and determining the star trace as a function of slit
position.  We then applied a wavelength solution to the
rectified images by identifying the OH sky lines with reference to a
list of vacuum wavelengths from the Kitt Peak National Observatory Fourier
Transform Spectrograph\footnote{Available at
  \url{http://www.astro.caltech.edu/mirror/keck/inst/nirspec/data/oh.lst}},
resulting in 2-D images rectified both spatially and spectrally.

For the NIRSPEC objects, we took four 900 s exposures of each galaxy
or galaxy pair, moving the object(s) along the slit for each
integration.  In order to subtract the sky background, we constructed
a sky frame from the temporally adjacent images; after scaling and
smoothing in the 
spatial direction, this sky frame was subtracted from the science
image.  Sky subtraction was done slightly differently for the ISAAC
observations, which were taken using ABBA offsets: a sky frame made from
the sum of the A images was subtracted from the B images, and vice versa.
Further background subtraction was done for both the NIRSPEC and ISAAC
observations by fitting a
polynomial in the spatial direction at each wavelength bin, avoiding
the positions of any bright objects on the slit; this removed some of
the residuals of the sky lines.  Finally, we produced a fully reduced,
two-dimensional spectrogram of each galaxy by registering and averaging
the individual frames, excluding bad pixels identified from combined dark
and flat-field images.  This step also produced a two-dimensional
frame of the statistical 1 $\sigma$ error appropriate to each pixel.
The last step was to extract one-dimensional spectra of each galaxy;
this was done by summing the pixels containing a signal along the
slit.  The same aperture was then used to extract a variance spectrum
from the square of the error image described above; the square root of
this is a 1 $\sigma$ error spectrum which was used to
determine the uncertainties in the line fluxes and widths.

\subsection{Flux Calibration}
\label{sec:fc}

In order to put the one-dimensional spectra onto an absolute flux
scale, we observed A0 and A2 stars from the list of UKIRT photometric
standards\footnote{Available at
\url{http://www.jach.hawaii.edu/JACpublic/UKIRT/astronomy/calib/}}.
These typically have $K \simeq 7$ mag,
and were observed at similar airmass and with the same instrumental
configuration as the galaxies themselves.  Flux calibration was done by scaling
the spectral energy distribution of Vega \citep*{cbc96} according to
the magnitude of the standard used, and dividing the spectrum of the
standard star by this scaled Vega spectrum.  This gives a sensitivity
function in counts per unit flux density, by which we divided our
one-dimensional galaxy spectra.  Because the spectra of A stars are
relatively smooth at the wavelengths of interest, they provide a
measurement of the atmospheric absorption, and dividing our galaxy
spectra by the sensitivity function therefore corrects for atmospheric
absorption. 

The uncertainties in the flux calibration process are both substantial and
difficult to quantify; however, we have attempted to estimate them in
several ways.  As described above, we extracted 1 $\sigma$ error
spectra for each of the galaxies; these primarily reflect the
noise of the sky background.  By integrating the flux in the variance
($\sigma^2$)
spectrum at the position of $\Ha$ and taking the square root of the
result, we can measure the random error
associated with the observation; this is $\leq$10\%.  More difficult
to measure are systematic errors: the largest sources of uncertainty are
the flux lost due to imperfect centering of the objects on the slit,
seeing and seeing variations, and the possibility of the objects being
larger than the slit itself.
We can get a sense of the importance of these effects by comparing the fluxes
received in each of the individual exposures which were co-added to
produce our final spectra.  We find that flux levels between exposures
vary by about 30\% (1 $\sigma$); this includes random as well as
systematic error.  The uncertainty in the mean flux of
our three or four exposures is then 15-20\%. This accounts for variations in
object centering and seeing, but not for flux consistently lost due to
the width of the slit.  As the galaxies observed are small
($r_{1/2}\sim$ 0\farcs2--0\farcs3 \space in an $HST$ WFPC2 pointing which
includes several of them), we assume that in most cases the flux loss is not
significant; however, a few of the galaxies are particularly irregular
and extended, and in these cases the flux
loss may be significant.  We can perform a further check by
calibrating the same object with several different standard stars; in
doing so we find variations in flux of 15\% at maximum, and usually
much less (again, 1 $\sigma$).  Because we have $K^{\prime}$-band photometry
for one of the galaxies in our sample (Q1700-BX691, one of the few in
which we detect a continuum signal), we can compare the
photometric flux with the continuum flux; we find that our spectrum
underestimates the photometric flux by a factor of 1.3, or about
25\%.  Because the continuum is so faint, this measurement is subject
to large errors, and is more a test of our sky subtraction than of our
spectrophotometry.  
We have also extracted one-dimensional
spectra of the standard stars with a variety of aperture widths in
order to determine whether an aperture correction might be necessary;
we find that less than 5\% of the flux is lost with the apertures
used to extract the galaxy spectra.  As this is much smaller than other
sources of error, no aperture correction was applied. Based on all of
these tests, we take our measured fluxes as uncertain by about 25\%.
This uncertainty propagates directly into the derived luminosities and
star formation rates, and will be adopted in the analyses that follow.

\section{Comments on Individual Objects}
\label{sec:ind}

While our selection process naturally leads to a wide range of UV
properties, with $\cal R$ ranging between 23.1 and 25.5 mag (i.e.\ a factor
of 9 in luminosity), we see less variation in the $\Ha$ fluxes, which
vary only by a factor of 4.  The UV and $\Ha$ properties are not
necessarily correlated, however; some of the strongest $\Ha$
luminosity comes from the faintest UV objects.  Because we have
rest-frame UV spectra of the galaxies, we are confident that none of
them are AGN; we see no high-ionization emission lines, and few even show
Ly$\alpha$ emission.  The lack of strong [\ion{N}{2}] emission also
indicates that the galaxies are not AGN.  \citet{vo87} find that AGN
have $\log$ [\ion{N}{2}]$\lambda6583/\Ha \gtrsim -0.2$, while we detect
[\ion{N}{2}] emission in only one case, and that weakly.  We hesitate
to infer anything about the metallicity of the galaxies based on the
absence of these lines, however, given the limited S/N of our data.
The galaxies are faint, with no 
spectroscopically detected
continuum in most cases, and we have not yet obtained rest-frame
optical magnitudes; therefore we are unable to
calculate $\Ha$ equivalent widths.  We see a larger
variation in the velocity dispersion $\sigma$ than in previous samples
of comparable size at high redshift \citep{pss+01}, but the most
notable feature of our sample is the six galaxies which show evidence
of ordered rotation, as we discuss in \S~\ref{sec:rot}.  We comment on
each object below.

\textit{Q1623-BX376:}  This is one of the brightest rest-frame UV
objects in our sample, and the only one in which the star formation
rate calculated from the UV emission is unambiguously higher than that
from the $\Ha$ emission (see \S~\ref{sec:sfrs}).  In
ground-based imaging it appears extended, with a fainter component
extending $\sim2$\farcs 5 to the west.  The association between the
two components is less clear with higher-resolution imaging (see
Figure~\ref{fig:hst}); however, the $\Ha$ emission also consists of
two lines at the same redshift, separated by 2\farcs 5.  We have
extracted spectra for both components, as shown in
Figure~\ref{fig:specs}; the primary component is labeled Q1623-BX376a,
and the fainter Q1623-BX376b.  Because our optical photometry treated both
components as a single extended object, we sum the fluxes from both
lines in order to calculate the $\Ha$ star formation rate in
\S~\ref{sec:sfrs}. 

\textit{Q1623-BX428:}  Unfortunately this galaxy lies at a redshift
such that $\Ha$ falls very close to a strong sky line, to which we
have lost significant flux.  This can be seen clearly in
Figure~\ref{fig:specs}, where the sky line falls just to the left of
$\Ha$.  Because of the loss of flux we are able to place a lower
limit on the $\Ha$ star formation rate, but the sky subtraction has
affected the line profile such that the velocity dispersion cannot be
determined. 

\textit{Q1623-BX447:}  This is one of the six galaxies for which we derived
rotation curves from tilted $\Ha$ emission lines; it is also one of the few
for which we have \textit{HST} imaging, which shows it to be
morphologically complicated (see Figure~\ref{fig:hst}).  We also see from
the \textit{HST} image that our slit was offset from the most extended
axis of the galaxy by $\sim$ 60 degrees.

\textit{Q1623-BX511:}  Of the six galaxies for which we were able to
derive rotation curves, this has the smallest $\Ha$ flux and hence the
smallest spread in velocity and the largest uncertainties.  The $\Ha$
emission falls between two bright sky lines, as can be seen in
Figure~\ref{fig:specs}.  At $\cal R\mathrm{=25.37}$ it is among the
faintest UV objects in our sample as well. 

\textit{Q1700-BX691:}  This is the only galaxy in which we clearly
detect [\ion{N}{2}] $\lambda\lambda$6549, 6583 and [\ion{S}{2}]
$\lambda\lambda$6717, 6734 emission lines as well as 
$\Ha$.  All of the lines are tilted in the two-dimensional spectra,
providing strong evidence for rotation.  The $\Ha$ rotation curve
reaches a velocity of $\sim240$ \kms at $\sim9$ kpc, with no
sign of flattening; this is clearly a massive system.  The fact
that we see [\ion{N}{2}] and [\ion{S}{2}] lines suggests a relatively
high metallicity; however, we defer a calculation until we are able to
obtain measurements of [\ion{O}{3}] in the $H$-band.  Interestingly
this is among the faintest UV objects in our sample, with $\cal
R$=25.33.  A $K^\prime$-band image of this object (\citealt{tmm98};
private communication) shows it to be extremely red, with $\cal
R$$-K^\prime=5.10$.  The $K^\prime$-band image also shows that our
slit was fortuitously aligned with the major axis.  

\textit{Q1700-MD103:}  This galaxy has the strongest $\Ha$ emission in
our sample, and hence the largest $\Ha$-derived star formation rate, 27 $\msun$
yr$^{-1}$.  It is also one of the six objects in which we detect
rotation.

\textit{Westphal-BX600:}  One of the six objects in which we detect
rotation, this galaxy is second only to Q1700-BX691 in rotational
velocity and implied mass.  We detected $\Ha$ emission
serendipitously, while observing the nearby $z\sim3$ galaxy
Westphal-MD115.  This object had been previously classified as a $z\sim2$
galaxy candidate based on its rest-frame UV colors, but it has not yet been
observed with LRIS.  Although we have no optical redshift, we believe
the line detected here to be $\Ha$ because its UV colors are entirely
consistent with a redshift of $z=2.16$; the contamination fraction in
the optical color selection process is less than 10\%, with most of the
interlopers being galaxies at low redshift ($z=0.05$--0.15).  We do
not know of any strong emission lines which would fall in our spectral
window for a galaxy in this redshift range;  for a redshift of
$z=0.008$, \ion{He}{1} (2.058 \micron; \citealt{lgl+01}) would fall at
the wavelength of the observed line, but then we would also expect to
see stronger Br$\gamma$ (2.166 \micron) emission at 2.18 \micron,
which we do not. 

\textit{SSA22a-MD41:}  This is one of the three galaxies which were
observed with the ISAAC spectrograph on the VLT.  Conditions were not
photometric during the ISAAC run, so we are only able to place a lower
limit on the $\Ha$ star formation rate.  We detect rotation in the
$\Ha$ emission, with a large spatial extent of nearly $\pm10$ kpc.

\textit{Q0201-B13} and \textit{CDFb-BN88:}  The other two galaxies
observed with ISAAC. As with SSA22a-MD41, we place lower limits on the
star formation rate from $\Ha$.  Q0201-B13 shows some evidence of
rotation in a slight tilt of the emission line, but the
signal-to-noise ratio (S/N) is too low to construct a reasonable
rotation curve.

\textit{Q1623-BX432, Q1623-BX449, Q1623-BX522, Q1623-MD107,
  Q1700-BX717, Q1700-MD109:} These are the remaining
  objects in the sample.  They span a factor of three in $\Ha$
  luminosity, from Q1623-BX432 at the bright end to Q1623-BX449 at the
  faint end, but none show evidence of velocity shear.  Our only
  kinematic information about these objects comes from the velocity
  dispersion; for three of the fainter objects (Q1623-BX449,
  Q1623-MD107, and Q1700-BX717) we were only able to place an upper
  limit on this quantity.

\textit{Non-detections:}  There are 10 galaxies which we observed with
NIRSPEC but failed to detect.  Four of these are accounted for by two
observations in which we did not detect either of the galaxies we
placed on the slit; in the cases of the other six, we detected one of
the galaxies on the slit, but missed the other.  For one of these the
optical redshift was unknown, so our hopes for detecting it were not high.
These 10 non-detections could have a variety of explanations, including
errors in our optical redshifts (which are of marginal quality in many
cases), in the astrometry, or in the guiding and tracking of the
instrument and telescope.  The objects could also be intrinsically
faint due to extinction or a decline in the star formation rate, as
discussed in \S~\ref{sec:sfrs}.

\begin{inlinefigure}
\centerline{\epsfxsize=8cm\epsffile{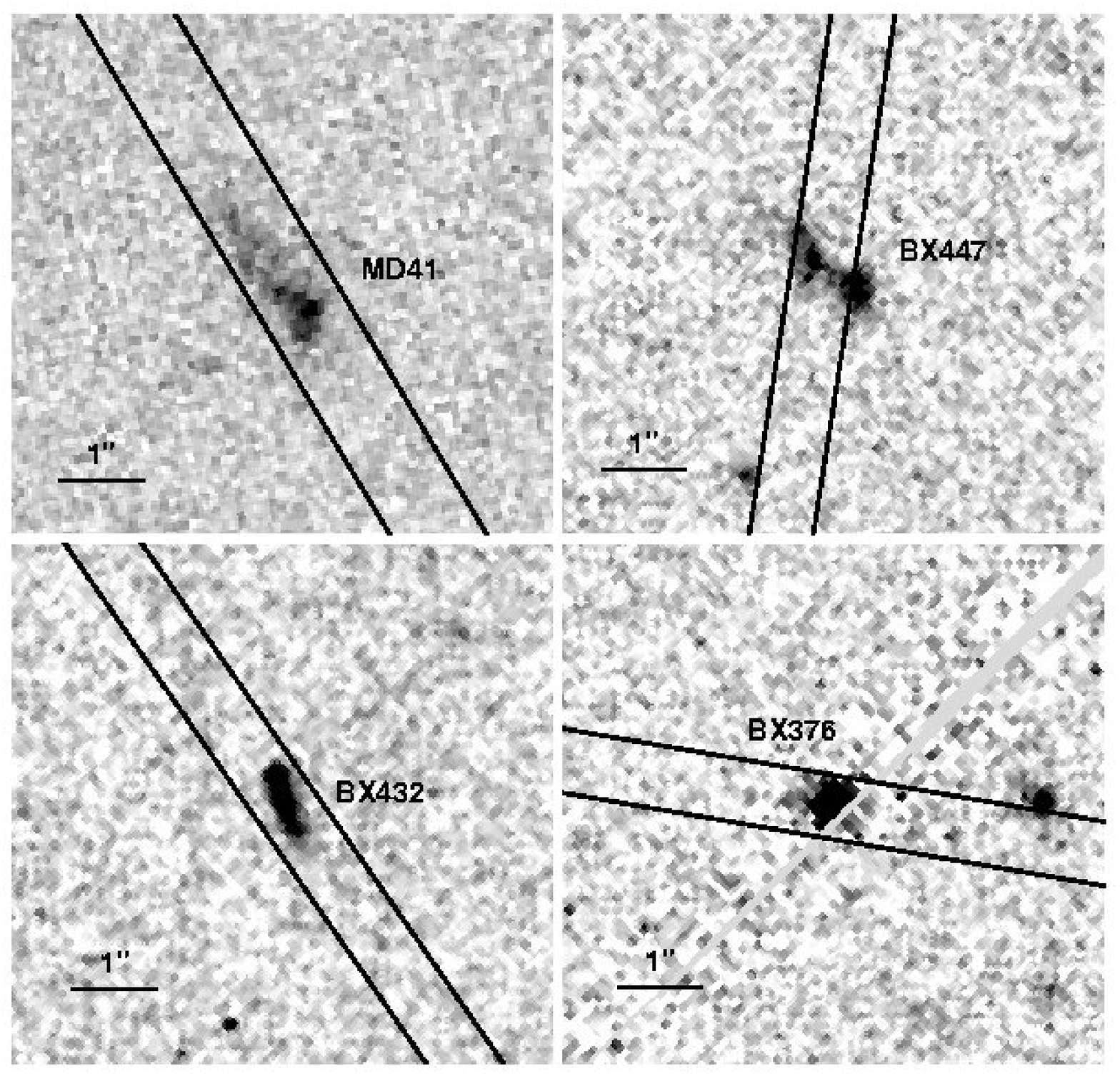}}
\figcaption{\textit{HST} WFPC2 images of four of the galaxies in our
  sample.  North is up and east to the left in all images, and
  positions of the slit are marked.  \textit{Upper left:}
  SSA22a-MD41, one of the objects in which we detect rotation, and the
  only one in which the slit was intentionally aligned with the major
  axis.  \textit{Upper right:} Q1623-BX447, another of the objects
  which show evidence of velocity shear.  In this case the slit and
  the major axis were misaligned by $\sim60$ degrees.  \textit{Lower
  left:}  We detect strong $\Ha$ emission from Q1623-BX432, but saw no
  evidence of velocity shear despite the near alignment of the slit along
  the major axis.  \textit{Lower right:} Q1623-BX376, the object with
  the largest velocity dispersion in our sample.  We also detect
  $\Ha$ emission at the same redshift from the object on the right;
  the two were classified as one extended object in our ground-based
  photometry.  The gray line running through the image is the boundary
  between two of the wide-field detectors.
\label{fig:hst}
}
\end{inlinefigure}

\section{Kinematics}
\label{sec:kin}

\subsection{Rotation}
\label{sec:rot}

\begin{figure*}
\centerline{\epsfxsize=18cm\epsffile{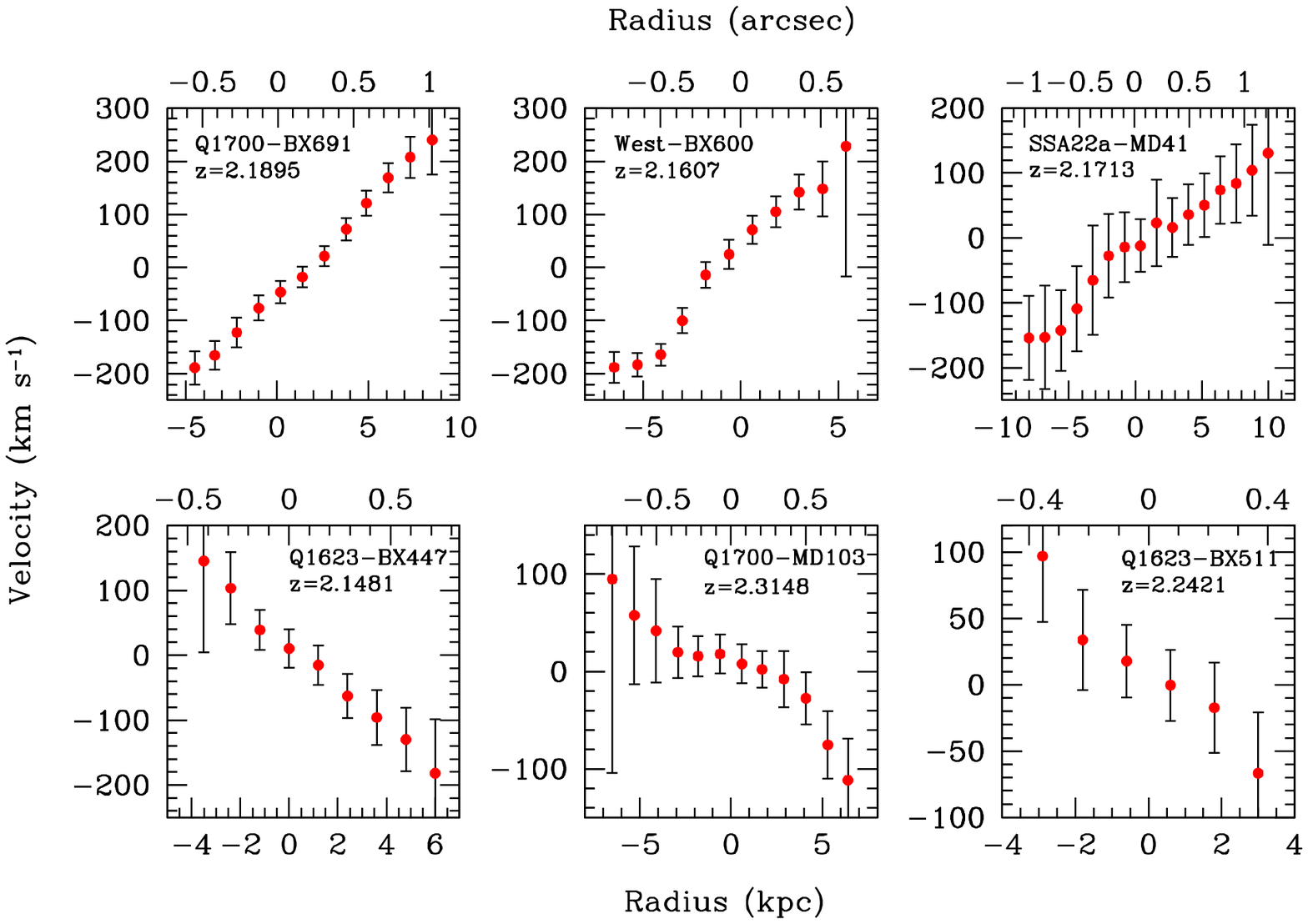}}
\figcaption{Rotation curves for the six galaxies that show spatially
  resolved, tilted $\Ha$ emission lines.  From the rotational
  velocities and radii we derive lower limits on the mass of each
  galaxy; the mean dynamical mass is $\langle M_{dyn}\rangle \geq 4
  \times 10^{10} \msun$.  The galaxies are shown in order of
  decreasing mass; from upper left, Q1700-BX691, West-BX600,
  SSA22a-MD41, Q1623-BX447, Q1700-MD103, and Q1623-BX511.  Note that
  the points are correlated due to the seeing of $\sim0.5$\arcsec.  We have
  used a cosmology with $H_0=70\;{\rm km}\;{\rm s}^{-1}\;{\rm Mpc}^{-1}$,
$\Omega_m=0.3$, and $\Omega_{\Lambda}=0.7$ for the transformations
  between arcsec and kpc.
\label{fig:curves}
}
\end{figure*}

Six galaxies in our sample of 16 show evidence of velocity shear, in
the form of a spatially resolved, tilted $\Ha$ emission line.  We have
constructed rotation curves for these objects by fitting a Gaussian
profile in wavelength to the emission line at each spatial location
along the slit, summing three pixels in the spatial direction at each
point in order to approximate the seeing of $\sim0$\farcs5.  Velocity
offsets were measured with respect to the systemic redshift of the
galaxy as determined from the central wavelength of the integrated $\Ha$
emission line; when
possible, the spatial center was defined by summing the spectra in the
dispersion direction without including the emission line and locating the
center of the continuum.  For those with no apparent continuum
emission (Q1700-MD103 and Q1623-BX511), the center was defined as 
the spatial center of the emission line.  The 2-D emission lines are
shown in Figure~\ref{fig:2dlines} and the rotation curves in
Figure~\ref{fig:curves}.  The observed velocities range from $\sim50$ to
$\sim240$ \kms, comparable to those observed in local galaxies
and up to $z\sim1$ \citep{vfp+96,vpf+97}.  In most cases
they show no sign of flattening at a terminal velocity; the
blue-shifted end of the curve of West-BX600 is the only one that
appears to flatten, and this is probably caused by imperfect
subtraction of an adjacent sky line.

There are several systematic effects to be considered here; most of
them result in an underestimation of the rotational velocity.  Except
in the case of SSA22a-MD41, no attempt was made to align the slit with
the major axis of the galaxy (position angles were chosen in order to
place two objects on the slit; see \S~\ref{sec:obs}); in fact in most cases our
ground-based images do not have sufficient resolution to allow the
determination of a major axis.  In the $K^{\prime}$-band image of
Q1700-BX691, however, it appears that here our slit was
fortuitously aligned with the major axis of the galaxy.  We also have
an \textit{HST} WFPC2 image of Q1623-BX447 (see Figure~\ref{fig:hst}) in
which it is apparent that the position angles of the slit and the
galaxy differ by $\sim60$ degrees.  In the other three cases, the slit
and the major axis were misaligned by an unknown 
amount.  In addition the inclinations of the galaxies are not known.
Given a random inclination and a random slit orientation, we will on
average underestimate the rotational velocity by a factor of
$(\pi/2)^2\simeq2.5$, where a factor of $\pi/2$ (the inverse of the
average value of $\sin x$ over the interval $(0,\pi/2)$) comes from
each effect.  Also, 
because all or most of 
each galaxy falls within the slit, the velocity we measure at each
spatial point along the slit is biased away from the maximum projected
velocity at the major axis by the lower velocities of points away from
the major axis.  We must also consider the possibilities of uneven
distribution of $\Ha$ emission and non-circular motions; both of these
are likely, given the irregular morphologies of the galaxies (see
Figure~\ref{fig:hst}).  A concentration of $\Ha$ away from the major
axis of the galaxy would lead to an underestimate of the rotational
velocity, but the effect of non-circular motions is more difficult to
predict. Typically many of these effects are modeled and corrected for
in rotation curves for less distant galaxies
\citep{vfp+96,vpf+97,smbb02}.  Given the chaotic, or unknown,
morphologies in our sample, we have not attempted to model these
corrections.

\begin{inlinefigure}
\centerline{\epsfxsize=9cm\epsffile{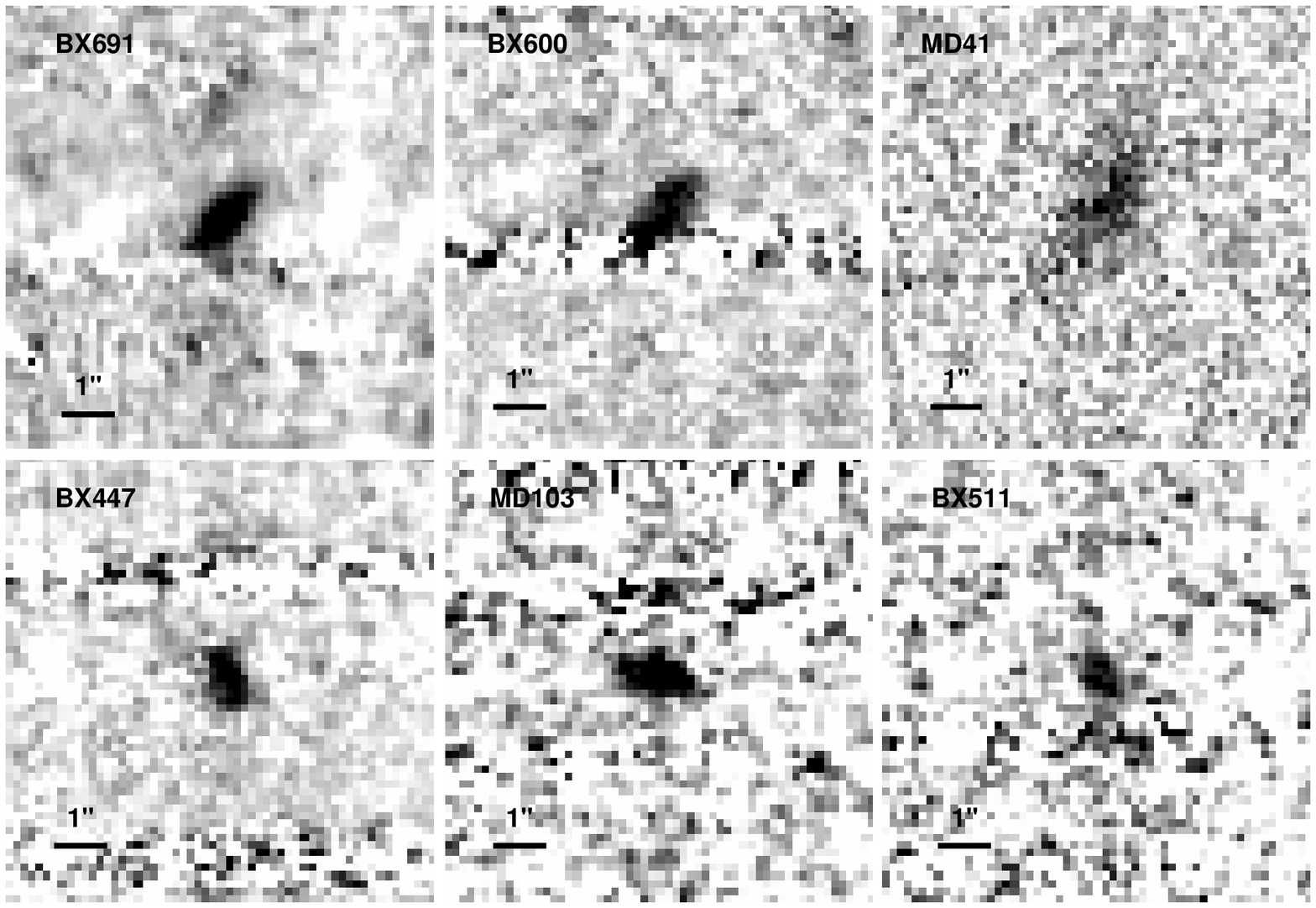}}
\figcaption{The two-dimensional spectra of the galaxies for which we have
  derived rotation curves, showing the tilt in the $\Ha$ emission
  line.  From upper left, the galaxies are Q1700-BX691 at $z=2.1895$;
  West-BX600 at $z=2.1607$; SSA22a-MD41 at $z=2.1713$; Q1623-BX447 at
  $z=2.1481$; Q1700-MD103 at $z=2.3148$; and Q1623-BX511 at
  $z=2.2421$.  A tilted [\ion{N}{2}]$\lambda 6584$  emission line is
  visible above $\Ha$ in the spectrum of Q1700-BX691.  The $x$ axis is
  spatial, with 1\arcsec \space scale bars shown, and $y$ is the
  dispersion direction. 
\label{fig:2dlines}
}
\end{inlinefigure}

We have used archival \textit{HST} WFPC2 images that contain two of
these galaxies, SSA22a-MD41 (in the F814W filter; proposal ID 5996)
and Q1623-BX447 (F702W; proposal ID 6557).  We reduced the images
following the drizzling procedure outlined in the \textit{HST} Dither
Handbook \citep{k02}; see \citet{fh02} for more details.  The images
are shown in Figure~\ref{fig:hst}, with the position of the slit marked.
Neither appears to be a well-formed disk; most of the rest-frame UV emission in
SSA22a-MD41 is concentrated in a knot at the southwest edge, and
Q1623-BX447 shows two distinct areas of emission.  It is interesting
to contrast these with images of two other galaxies for which we did
not detect rotation:  Figure~\ref{fig:hst} also shows images of
Q1623-BX432 and Q1623-BX376, which are also contained in the Q1623
pointing and which also appear irregular.  This demonstrates the
difficulty of predicting the kinematics of these objects from even
high-resolution imaging; complicated morphologies make inclinations
and major axes difficult to determine, and objects with similar UV
continuum morphologies may exhibit quite different $\Ha$ kinematic
properties.  We also point out that 
the $\Ha$ and UV emission may not be coincident; \citet{pss+01}
observed nebular line emission extending $\sim1$\arcsec \space beyond
the UV emission in a galaxy at $z=3.2$, and similar effects have been
seen in local galaxies \citep{lvc+96,dshl98,jlvc00}.  Specifically,
\citet{cgc+00} compared $\Ha$ and UV emission in six nearby starburst
galaxies, finding that the $\Ha$ and UV fluxes were well correlated in
three of the systems, but that they showed different morphologies in
the other three. 

Although we have no direct evidence that these galaxies are in fact
disks, we make this assumption in order to use the
radius $r$ and the circular velocity $v_c$ to calculate the enclosed
mass, 
\begin{equation}
M_{dyn}=v_c^2r/G.
\end{equation}
Since we have neither well-defined terminal
velocities nor spatial centers for these objects, we have calculated
lower limits on the masses by using half of the total spread in both
velocity and distance, for $v_c$ and $r$ respectively.  We obtain an
average dynamical mass of $\langle M \rangle \geq 4 \times 10^{10}
\msun$; individual masses for each galaxy 
are shown in Table~\ref{tab:kin}.  As the $\Ha$ emission
traces only the central star-forming regions of these objects, which
are probably baryon-dominated, the masses derived are
underestimates of the total halo masses of the galaxies.  We can use an
order of magnitude argument to estimate the total masses:  for
$\Omega_b=0.02 h^{-2}$ and $\Omega_m=0.3$, $\Omega_m/\Omega_b\sim7$,
and the universe contains about six times more dark than baryonic matter.  We
therefore expect the total masses of the galaxies to be about seven
times larger than their stellar masses, and we place a lower limit of
$M \gtrsim 3 \times 10^{11} \msun$ 
on the typical halo mass of these galaxies.  This is generally consistent with
mass estimates from the clustering properties of LBGs at $z\sim3$:
\citet{asg+98} find a typical mass of $8 \times 10^{11} h^{-1}
\msun$ for a $\Lambda$CDM model, based on the number density and
correlation length of the galaxies.  Other
analyses yield similar results \citep{bcf+98,gd01}.  We defer an
analysis of the clustering of the $z\sim2$ galaxies to a
later work.  We can also compare our mean baryonic mass with the median
stellar mass from population synthesis models found for LBGs at
$z\sim3$ by \citet{ssa+01}, $m_{\rm{star}}=1.2 \times 10^{10} h^{-2}
\msun$; again the two are in rough agreement.

There are few other examples of such rotation curves at redshifts of
$z \gg 1$.  \citet{lcp+02} have recently reported a rotation curve of a
gravitationally lensed galaxy at $z=1.9$;  the rotation curve looks
much like those we present here, with $v\gtrsim200$ \kms at a radius
of $\sim1$\arcsec, although when the lensing correction is applied
this radius corresponds to only $\sim1$ kpc.  \citet{gbt+02} have used
millimeter interferometry to observe rest-frame 335~\micron \space continuum
and CO(3--2) line emission from a massive submillimeter galaxy at $z=2.8$;
their data indicate a rotating disk with velocity $\geq420$ \kms at
$\sim8$ kpc in radius.  From observations of [\ion{O}{3}] at
$z\simeq3.2$, \citet{mvc+03} present a rotation curve with a velocity
of 108 \kms at $\simeq12$ kpc.  Also in observations of [\ion{O}{3}]
and H$\beta$ in 15 LBGs 
at $z\sim3$, \citet{pss+01} see two spatially resolved and tilted emission
lines, but the observed velocities reach only $\sim50$ \kms.  Simply
counting the instances of rotation shows that the two samples are different
at the 95\% confidence level; the difference is actually more
significant, because this test does not account for
the larger rotational velocities at $z\sim2$.  It is
interesting that we see stronger evidence for rotation in a
sample of similar size at $z\sim2$, and we will spend a moment
speculating on the possible reasons for this.  Poorer seeing during
the $z\sim3$ observations could perhaps account for the differences;
this does not explain the larger values of the velocity dispersion
$\sigma$ we see at $z\sim2$ (see \S~\ref{sec:veld}), however, as these
should be unaffected even if the lines are spatially unresolved.  It
might then be that $\Ha$ is a more sensitive probe of rotation and
velocity dispersion than [\ion{O}{3}] because of higher surface
brightness; but \citet{pss+01} typically measured
[\ion{O}{3}]$\lambda5007/\Hb \sim 3$, and $\Ha/\Hb \sim3$ as well, so
$\Ha$ and [\ion{O}{3}]$\lambda5007$ should have roughly comparable
strengths in the $z\sim3$ galaxies.  We also note that in rotation
curves for which they had both $\Ha$ and [\ion{O}{3}]$\lambda5007$
data, \citet{vfp+96} found that the flux distributions and velocities
of the two lines matched well.  \citet{lcp+02} also have both $\Ha$
and [\ion{O}{3}]$\lambda5007$ observations for their rotation curve,
and again the two lines give comparable results.  The differences could also
be due to S/N effects; but the $z\sim3$ galaxies were generally
observed with longer integration times than those in the current
sample, and their spectra have S/N comparable to or higher than
that of those presented here.  We should also discuss the possibility
that we may be 
observing different populations of galaxies at $z\sim2$ and $z\sim3$.  We
therefore consider the evidence for other intrinsic differences between
galaxies at the two redshifts.  The most obvious of these is apparent UV
luminosity; the galaxies of \citet{pss+01} are brighter than those
presented here, with only a few exceptions.  This is simply because
the brightest galaxies were selected for IR observation at $z\sim3$,
but not at $z\sim2$.  As discussed in \S~\ref{sec:obs}, however, the $z\sim2$
selection criteria were chosen so that the galaxies they select would
have SEDs similar to galaxies at $z\sim3$.  If we are indeed looking
at different sets of objects at $z\sim2$ and $z\sim3$, both the
average and range of their far-UV properties must be similar (although
we do sample the luminosity function more deeply at $z\sim2$).  It is
also possible that we are observing the two samples to different
radii:  surface brightness is a strong function of redshift, scaling
as $(1+z)^4$, and this may limit the radii to which we can observe the
galaxies at higher redshift.  Star formation progressing to larger
radii in the disks at later times could produce a similar effect.
It is also possible that our stronger evidence for
rotation reflects an increase in the number of rotating galaxies and
their rotational speeds between $z\sim3$ and $z\sim2$.  With the
present data such a conclusion would be premature, however, since we
cannot rule out all observational effects.

It is interesting to consider objects such as these in the context of
hierarchical models of galaxy formation.  We compare our data with
predictions of the properties of LBGs at $z\sim3$
\citep*{mmw98,mmw99}, although it is not yet clear how the current
sample and the $z\sim3$ galaxies are related.  LBGs are thought to be
the central galaxies of the most massive dark halos present at
$z\sim3$, and they are predicted to be small and to have moderately high
halo circular velocities but low stellar velocity dispersions.  For a
$\Lambda$CDM cosmology, \citet{mmw99} predict that the median
effective radius $R_{\rm eff}$ (defined as the semimajor axis of the
isophote containing half of the star formation activity) is about 2
$h^{-1}$ kpc, and most galaxies should have $R_{\rm eff}$ between 0.8
and 5 $h^{-1}$ kpc.  While the maximum radial extent of some of our
rotation curves is larger than this, it is likely that the galaxies
are visible at radii beyond $R_{\rm eff}$, and these predictions are
consistent with our measurements of half-light radii from the WFPC2
images.  \citet{mmw99} also predict a median halo circular velocity of
290 \kms for $\Lambda$CDM, with most galaxies falling between 220 and
400 \kms, and a median stellar velocity dispersion of $\sim120$ \kms.
Both of these predictions are reasonably consistent with our data,
considering that we have not corrected our circular velocities for
inclination or slit alignment effects, and that our velocities are
lower limits due to the lack of flattening in the rotation curves.  In
fact, as noted above, the 
$z\sim2$ galaxies are a better match to these predictions than the
$z\sim3$ LBGs, which have observed rotational velocities of only
$\sim50$--100 \kms and velocity dispersions of $\sim70$ \kms.

Finally, additional observations will clarify the
kinematics of the $z\sim2$ sample.  High resolution imaging in both the
optical and the IR will allow a determination of the morphologies of the
galaxies and the extent of the rest-frame optical emission;
spectroscopic observations with varying position angles will provide 
strong constraints on rotating disk models.  We are also optimistic
about the possibility of obtaining a larger sample of rotation curves,
since those presented here represent almost 40\% of the galaxies
observed.  Looking farther into the future, integral field IR
spectrographs that provide kinematic information at high spatial
resolution over a contiguous region encompassing the entire galaxy
will be ideal for probing the dynamics of high redshift galaxies; this
may be the only way that the kinematic major axes of these objects
can be determined.

\subsection{Velocity Dispersions}
\label{sec:veld}

We can obtain a limited amount of information about the dynamics and
masses of the galaxies by simply measuring the widths of the emission
lines.  We have measured the one-dimensional velocity dispersion
$\sigma$ by fitting a Gaussian profile to each emission line,
measuring its FWHM, and subtracting the instrumental broadening in
quadrature from the FWHM.  The instrumental broadening was measured
from the widths of sky lines, and is $\sim15$ \AA \space for NIRSPEC
and $\sim6$ \AA \space for ISAAC.  The velocity dispersion is then the
corrected FWHM divided by 2.355.   We find a mean velocity
dispersion of $\langle \sigma \rangle \sim110$ \kms, with a
maximum of 260 \kms.  The dispersions for each galaxy are shown
in Table~\ref{tab:kin}, with 1 $\Delta_{\sigma}$ uncertainties from
propagating 
the errors in each Gaussian fit (to avoid confusion stemming from
overuse of the the symbol $\sigma$, we use $\Delta_{\sigma}$ to represent
the standard deviation in the velocity dispersion).  Most of the lines are resolved; for
those that are not we have set an upper limit of 2 $\Delta_{\sigma}$.  Our
average velocity dispersion is $\sim60$\% higher than that
found from the widths of [\ion{O}{3}]$\lambda$5007 and H$\beta$ at $z\sim3$ by
\citet{pss+01}, who found a median of $\sim70$ \kms. \footnote{We also
  find a mean of $\sim70$ \kms in the [\ion{O}{3}]$\lambda$5007
  velocity dispersions of 
  a sample of 11 LBGs at $z\sim3$ which we observed with NIRSPEC in April 2001.
  These data are unpublished, and will be described in detail in a
  later work.}

Assuming that these velocities are due to motion of the gas in the
gravitational potential of the galaxy, we
can estimate the masses of the galaxies.  For the simplified case of a
uniform sphere,
\begin{equation}
M_{vir}=5\sigma^2(r_{1/2}/G).
\end{equation}
From the \textit{HST} image of the galaxies in the Q1623 field, we find
$r_{1/2}\sim0\farcs2$, which in our adopted cosmology corresponds to
$\sim1.6$ kpc at $z=2.3$.  We use this value to calculate the
masses shown in Table~\ref{tab:kin}.  Accounting for the lower limits
on four of the objects by using ASURV Rev.~1.2 \citep*{lif92}, a
software package which calculates the statistical properties of
samples containing limits or non-detections (survival analysis;
\citealt*{fn85}), we find a mean mass of $\sim 2\times10^{10} \msun$;
this is in general agreement with the rotationally-derived masses in
\S~\ref{sec:rot}. As we noted when deriving masses from the rotation
curves above, because the nebular emission comes mostly from the central
star-forming regions of high-surface brightness, the velocity
dispersions probably do not reflect the full gravitational potential
of the galaxies. 

There are several issues to consider in the interpretation of these
mass estimates.  In
addition to the obvious caveats related to the assumption of spherical
geometry, the uncertain value of $r_{1/2}$, and the sometimes large
uncertainties in $\sigma$, we should consider
whether or not the line broadening is indeed gravitational in origin.
Galaxy-scale starburst-driven outflows with speeds of several hundred
\kms have been shown to be ubiquitous in star-forming galaxies at
$z\sim3$ \citep{pss+01}.  These are measured from the offsets of
Ly$\alpha$ and the interstellar absorption lines with respect to the
nebular emission lines taken to define the systemic velocity of the
galaxy; Ly$\alpha$ is consistently redshifted with respect to the
systemic velocity, while the interstellar lines are blueshifted.  We
are unable to determine conclusively whether or not similar outflows
exist in the present sample, since in many cases the S/N ratios of our
rest-frame UV spectra are too low to determine redshifts from Ly$\alpha$ and
interstellar absorption lines with the necessary precision.  However,
for those objects that have spectra of sufficient quality, we have
measured the velocities of the interstellar absorption lines and
Ly$\alpha$ with respect to the $\Ha$ redshifts.  The results are shown
in Figure~\ref{fig:voff}.  We see that in this small sample, Ly$\alpha$ is
consistently redshifted by several hundred \kms, but that the
interstellar lines are both blueshifted and redshifted with respect to
$\Ha$.  This offers marginal support for the existence of outflows, but
clearly a larger sample is necessary.  Even if these outflows do 
exist, however, it is not clear that they would result in an increase
in the velocity dispersion.  Our velocity dispersions are from $\Ha$
emission, which we take to be coming primarily from nebular gas at the systemic
redshift of the galaxy, not from outflowing material.   In addition, a
correlation between the velocity dispersion and the speed of the
outflow (here defined as the average of $v_{\rm Ly \alpha}-v_{\rm
  neb}$ and $v_{\rm neb}-v_{\rm IS}$) might  be expected if the line
broadening were due to  
outflowing gas.  With this in mind we have
examined a sample of 23 galaxies at $z\sim3$ for which we have both velocity
dispersions from the width of the [\ion{O}{3}]$\lambda$5007 emission
line and outflow velocities from the offsets between the nebular,
interstellar absorption and Ly$\alpha$ redshifts.  We see no
evidence for a strong link between the velocity dispersion and the speed of the
outflow; the correlation coefficient between them is 0.13.  These
considerations lead us to believe that
the presence of outflows is not a strong argument
against gravitational broadening of the lines.
\begin{inlinefigure}
\centerline{\epsfxsize=9cm\epsffile{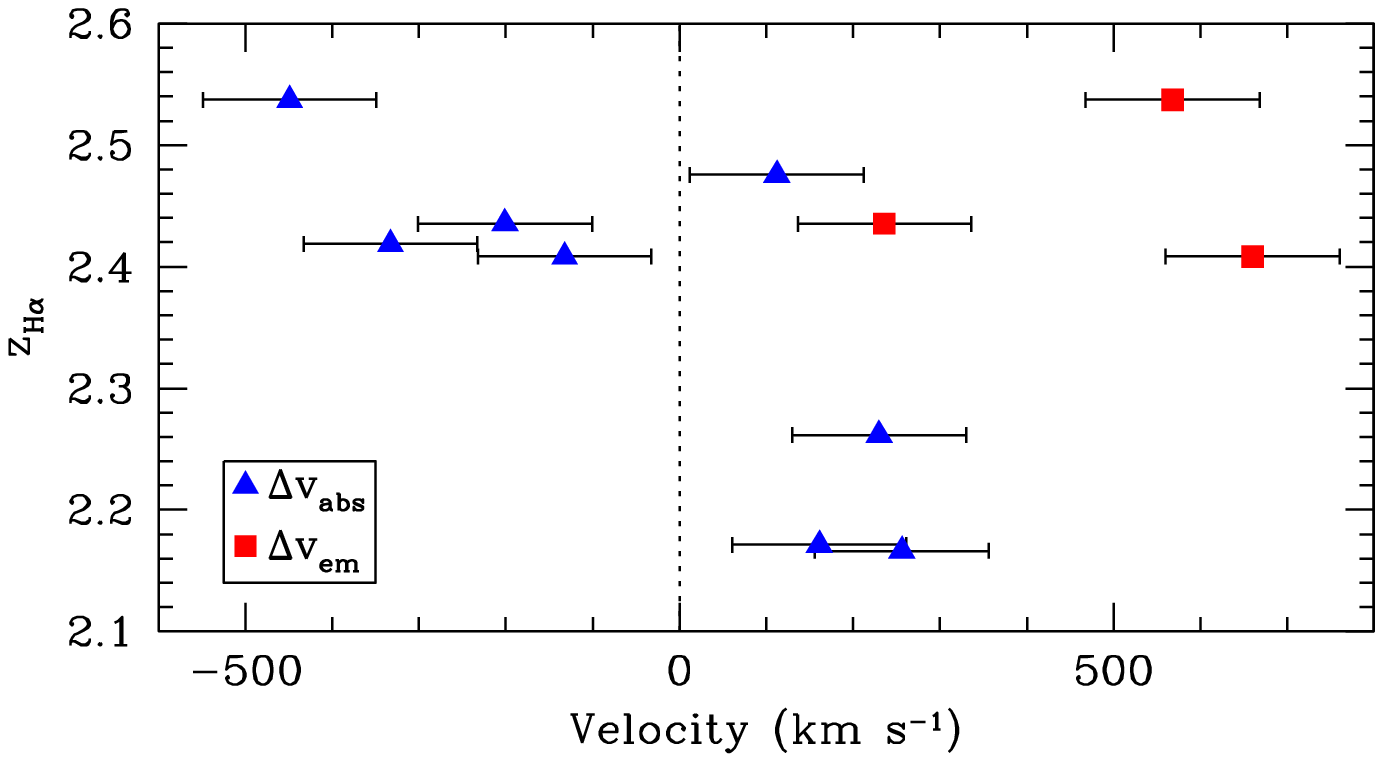}}
\figcaption{Velocity offsets between the systemic velocity of $\Ha$ and
the velocities of the interstellar absorption lines (blue triangles)
and Ly$\alpha$ emission (red squares).  The sample is small because
most of our galaxies do not have rest-frame UV spectra of sufficient
S/N to make this comparison.
\label{fig:voff}
}
\end{inlinefigure}
\begin{inlinefigure}
\centerline{\epsfxsize=9cm\epsffile{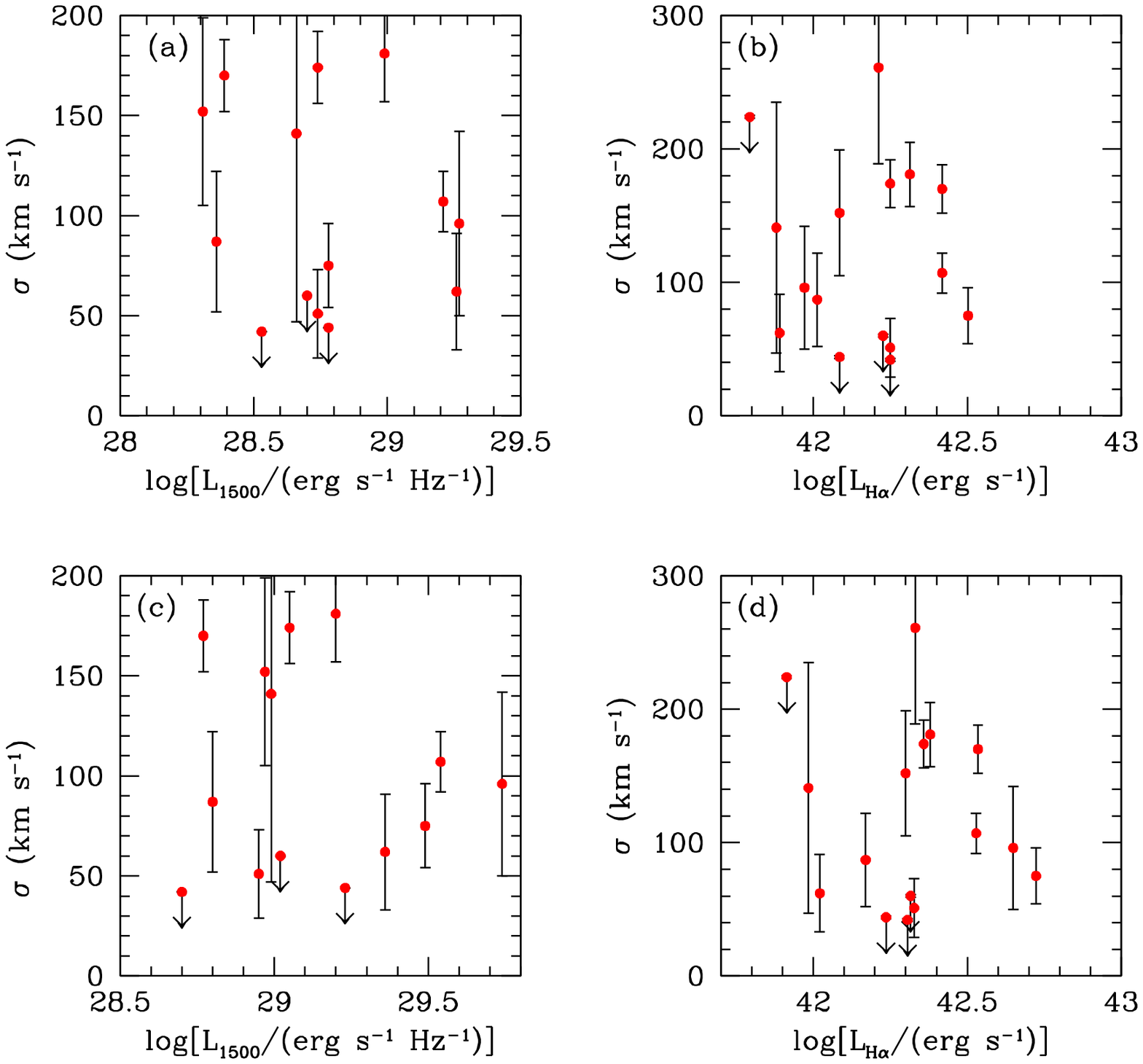}}
\figcaption{The velocity dispersion $\sigma$ plotted against the 1500 \AA
  \space continuum and  $\Ha$ luminosities, without extinction
  corrections (\textit{a} and \textit{b}), and corrected as described
  in the text (\textit{c} and \textit{d}).  Arrows indicate upper
  limits on $\sigma$.  See \S~\ref{sec:sfrs} for a discussion of the errors in
  luminosity.
\label{fig:sigl}
}
\end{inlinefigure}

We are also struck by the spatial complexity of some of these objects.
In particular, the $\Ha$ emission of Q1623-BX376 appears as two
lines at the same redshift but separated by 2\farcs5.  The brighter of
these, Q1623-BX376a, has the largest velocity dispersion in the
sample, and shows an asymmetric line profile (see
Figure~\ref{fig:specs}), with a blue-shifted tail extending about
0\farcs5 in the opposite direction from the fainter component,
Q1623-BX376b.  It is primarily this tail which is responsible for the
large velocity dispersion.  This faint emission is also visible in the
WFPC2 image shown in Figure~\ref{fig:hst} (where, unfortunately, the
galaxy falls on the border between two of the wide-field detectors).
Given the complicated structure of this object, we hesitate to
attribute its broad emission line purely to random gravitational
motions; galactic mergers or interactions could also produce such
broadened emission lines and disturbed morphologies. 

As a final test, we compare the one-dimensional velocity dispersion
with luminosity.  We see from Figure~\ref{fig:sigl} that neither the
1500 \AA \space continuum nor the $\Ha$ emission line luminosity
correlate with velocity dispersion, either with or without a correction for
extinction.  Such a lack of correlation is also seen for galaxies at $z\sim3$
\citep{pss+01}.  This does not necessarily mean that the line widths
are unrelated to the masses of the galaxies; it may be that large
variations in the mass-to-light ratio are blurring any trend.  We
conclude that while these caveats are important, none of them provide
a compelling argument against using the velocity dispersions to
estimate the masses of the galaxies; therefore for the moment we
will continue to do so. 

\section{Star Formation Rates and Extinction}
\label{sec:sfrs}

$\Ha$ emission is one of the primary diagnostics of the star formation
rate (SFR) in local galaxies, and therefore its observation at high redshift is
particularly valuable for the sake of comparison with nearby samples.
Redshifts of $z\lesssim2.6$ are the highest at which $\Ha$ can
currently be detected before it shifts out of the near-IR $K$-band
window.  Except for a few other observations of $\Ha$ at $z>2$
\citep{tmm98,kk00}, most determinations of the star formation
rate at high redshift have so far been based on the UV stellar
continuum and, to a lesser extent, the $\Hb$ emission line \citep{pss+01}.
Here we compare star formation rates for the 16 galaxies in our sample
deduced from the $\Ha$ flux and from the UV
continuum emission; as the two are affected differently by
dust and star formation history, our results can in principle tell us
about the extinction and stellar populations of the galaxies.  We have
calculated $\Ha$ SFRs following \citet{k98}: 

\begin{equation}
\textup{SFR } (\msun \textup{ yr}^{-1}) = 7.9 \times 10^{-42} \;L(\Ha) \:\:(\textup{erg s}^{-1})
\end{equation}

The nebular recombination lines are a direct probe of the young,
massive stellar population, since only the most massive and
short-lived stars ($M \gtrsim 10 \msun$) contribute significantly to
the ionizing flux.  Thus the emission lines provide a nearly
instantaneous measure of the SFR, independent of the star formation
history.  The above equation assumes a Salpeter IMF with upper and
lower mass cutoffs of 0.1 and 100 $\msun$ and case B recombination at
$T_e$=10,000 K.  It also assumes that all of the ionizing photons are
reprocessed into nebular lines, i.e.\ that they are not absorbed by dust
before they can ionize an atom, and that they do not escape the galaxy.

Ultraviolet-derived star formation rates were calculated from the
broadband optical photometry, using the $G$ magnitude as an
approximation for the 1500 \AA\ continuum (at $z = 2.3$, the
mean redshift of our sample, the central wavelength of the $G$ filter,
4830 \AA, falls at a rest wavelength of 1464 \AA).  SFRs were
calculated as follows \citep{k98}:

\begin{equation}
\textup{SFR } (\msun \textup{ yr}^{-1}) = 1.4 \times 10^{-28}\;L_{1500} \:\:(\textup{erg s}^{-1}\; \textup{Hz}^{-1})
\end{equation}

This relationship applies to galaxies with continuous star formation over time
scales of 10$^8$ years or longer; for a younger population, the UV
continuum luminosity is still increasing as the number of massive
stars increases, and the above equation will underestimate the star
formation rate.  The assumed IMF is the same as above. 

The fluxes and corresponding SFRs are summarized in Table~\ref{tab:sfr}, and a
comparison of the uncorrected star formation rates is shown in the
left panel of  Figure~\ref{fig:sfrs}.  The error bars reflect the
uncertainties in flux calibration of the $\Ha$ emission and the UV
photometry, about 25\% and 10\% respectively; for the $\Ha$ spectra
this includes both random and systematic error, as discussed in
\S~\ref{sec:fc}, and is likely an underestimate in the noisiest cases.
Uncertainties in the conversion from flux to SFR are not included.
There are four objects for which we are only able to place lower
limits on the SFR from $\Ha$: Q1623-BX428, in which the $\Ha$ line
fell on top of a strong sky line to which we have lost significant
flux, and SSA22a-MD41, Q0201-B13, and CDFb-BN88, which were observed
during non-photometric conditions (and calibrated with the least
extinguished exposure of a standard, in order to place lower limits).
Without correcting for extinction, we find $\rm{SFR_{\Ha} > SFR_{UV}}$
in all but five cases; four of these are the lower limits described
above.  We find $\langle\rm{SFR_{\Ha}/SFR_{UV}}\rangle = 2.4$; this was
computed using ASURV Rev.~1.2 \citep*{lif92}, a software package which
calculates the statistical properties of samples containing limits or
non-detections (survival analysis; \citealt*{fn85}).  This result is
in qualitative agreement with  previous observations of galaxies at $z
\gtrsim 1$:  \citet{ymf+99} find that the global star formation rate
derived from $\Ha$ exceeds 
that from the UV by a factor of $\sim3$, and \citet{hcs00} obtain
a measurement of SFR density from $\Ha$ at $0.7\leq z \leq1.8$ that is
a factor of 2--3 
greater than that estimated from UV data.  \citet{gbe+99} study a
sample of 13 galaxies at $z\sim1$ from the Canada France Redshift
Survey (CFRS); when the same \citet{k98} calibrations are used, their
data give an $\Ha$ SFR 1.9 times higher than the UV SFR, without
applying an extinction correction
\citep{ymf+99}.\footnote{\citet{ymf+99} and
  \citet{gbe+99} assume $H_0=50\;{\rm km}\;{\rm s}^{-1}\;{\rm Mpc}^{-1}$ and
  $q_0=0.5$. Using this cosmology lowers our SFRs by 5--10\%; the
  ratios of the $\Ha$ and UV rates are, of course, unaffected.}  It is
also comparable to the results of \citet{bk01}, who find
$\langle \rm{SFR_{\Ha}/SFR_{UV}\rangle = 1.5}$ for galaxies with SFR
$\gtrsim 1 \msun$ yr$^{-1}$ in a sample of 50 nearby star-forming
galaxies.  There is clearly a trend for the $\Ha$-derived SFRs to be
higher than those from the UV luminosity, in spite of differing
selection criteria;  both the \citet{ymf+99} and \citet{hcs00}
samples were selected in the IR, while ours is UV-selected and the
\citet{bk01} sample is drawn from local galaxies observed by the
Ultraviolet Imaging Telescope (UIT).  We will discuss possible reasons
for this trend below.   We also note that the one remaining
object with a larger UV SFR, Q1623-BX376, is a somewhat unusual
case.  It is bright and extended in the UV, and the $\Ha$ emission
appears in two distinct lines at the same redshift but separated by
2\farcs5.  Since the UV photometry encompassed both components we have
added the flux from both lines to calculate the $\Ha$ SFR, but it is
clear from the WFPC2 image of Q1623-BX376 (Figure~\ref{fig:hst}) that
the fainter of the two components is largely off the edge of the slit;
therefore we have likely missed some of the $\Ha$ emission.

\begin{inlinefigure}
\centerline{\epsfxsize=9cm\epsffile{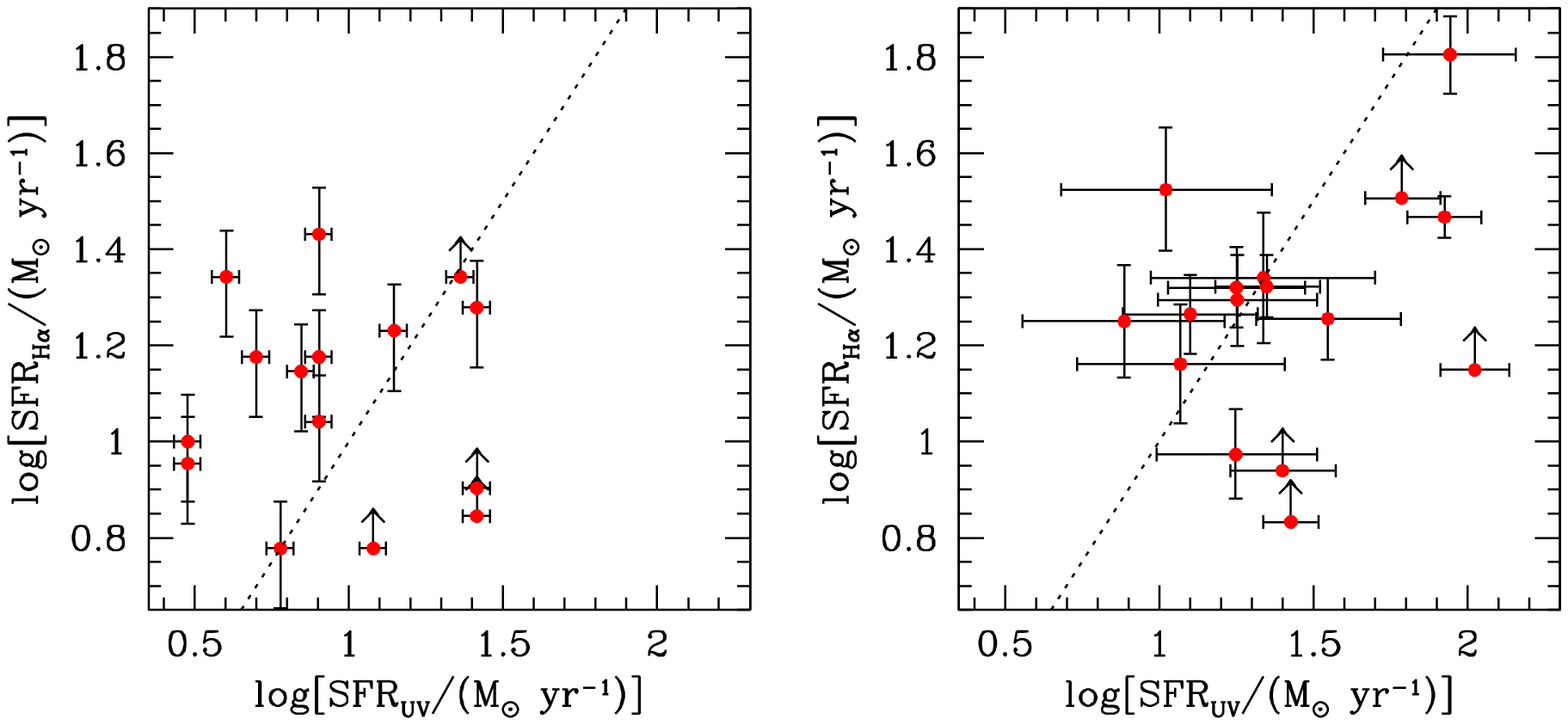}}
\figcaption{\textit{Left:} Star formation rates from $\Ha$ and UV emission,
  uncorrected for extinction.  Arrows indicate lower limits on
  the $\Ha$ SFR for objects observed during non-photometric
  conditions (SSA22a-MD41, Q0201-B13, and CDFb-BN88) or contaminated
  by sky lines (Q1623-BX428).  Errors are 25\% in SFR$_{\Ha}$ and 10\%
  in SFR$_{\mathrm{UV}}$,  reflecting uncertainties in flux calibration.
  Uncertainties in the conversion from flux to SFR are not included.
  \textit{Right:}  The SFRs corrected for extinction as described in
  \S~\ref{sec:sfrs}.  The error bars
  reflect uncertainties in $E(B-V)$ only; flux calibration errors and
  errors in conversion from flux to SFR are not included.  The dotted
  lines represent equal rates from $\Ha$ and UV emission.
\label{fig:sfrs}
}
\end{inlinefigure} 

There are at least two possible explanations for the larger $\Ha$
SFRs: dust extinction and the two star formation rate indicators' differing
sensitivities to the ages of stellar populations and star formation
histories.  Our observations are consistent with the assumption that
the ultraviolet emission generally suffers greater extinction than the
$\Ha$, as would be the case if 
both pass through the same clouds of dust.  However, in analogy to
local starbursts, it may be the
case that the UV and nebular line emission come from different
regions in the galaxies and encounter different amounts of dust
accordingly \citep{c97}.  In particular, it has been suggested
that the most massive stars are still embedded in the dust clouds in
which they formed, leading to greater extinction of the nebular line
emission.  This may be the case with Q1623-BX376, which is bright in the
rest-frame UV, with Ly$\alpha$ emission and strong interstellar
absorption lines, but which is undistinguished when observed in the
rest-frame optical.  

We can estimate the UV extinction using the observed broadband colors
and an assumed spectral energy distribution (SED); we have calculated
$E(B-V)$ in this way, using the $G-\cal R$ colors and an SED
corresponding to continuous star formation with an age of 320 Myr, the
median age found for LBGs at $z\sim3$ by \citet{ssa+01}.  Because
extinction corrections are highly sensitive to errors in color
measurements, we have made an effort to quantify the uncertainties and
biases in our photometry.  We added a large number of artificial
galaxies of known colors and magnitudes to the actual images, and then
recovered them using the same photometric tools that we applied to the
real data (see \citealt{a02,sas+03}).  We then selected artificial galaxies
whose recovered colors match our selection criteria, and sorted them
into bins by color and $\cal R$ magnitude.  We used these to measure
the mean and dispersion of $\Delta (G-{\cal R})=((G-{\cal R})_{\rm
  meas}-(G-{\cal R})_{\rm true})$, where the mean indicates systematic
biases in the recovered colors and the dispersion reflects the
characteristic measurement error, $\sigma (G-{\cal R})$.  For the
brightest galaxies in our sample (${\cal R} < 23.5$) both of these
quantities are small: $\langle \Delta (G-{\cal R}) \rangle \simeq
0.03$ and $\sigma (G-{\cal R}) \simeq 0.05$.  For those with ${\cal R} > 25$, 
we find $\langle \Delta (G-{\cal R}) \rangle \simeq 0.04$ and $\sigma
(G-{\cal R}) \simeq 0.14$.  For each galaxy in our sample, we have
used these statistics to correct the measured $G-{\cal R}$ color for
the bias, and the color error has been propagated to determine uncertainties in
$E(B-V)$; these range from 0.03 for the brightest galaxies to 0.08 for
the faintest.

After calculating $E(B-V)$ in this way, we used the \citet{cab+00}
extinction law to correct the $G$ magnitudes, and then used these to
recalculate the UV star formation rates.  For the sake of comparison
we have also corrected the $\Ha$ fluxes, assuming the same values of
$E(B-V)$; we found this to give better agreement between the corrected
UV and $\Ha$ SFRs than the \citet{c97} relation
$E_s(B-V)=(0.44\pm0.03)E_n(B-V)$ (where $E_s(B-V)$ is the color excess
of the stellar continuum and $E_n(B-V)$ is that of the nebular
emission lines).  There may be some justification for this: if indeed
there are galactic-scale outflows in these galaxies as in those at
$z\sim3$, then a screen of outflowing material may be obscuring all
regions equally.  Unfortunately we have no way of independently
measuring the nebular extinction with our current data, as we do not
have $H$-band measurements of $\Hb$.  It should also be noted that the
uncertainties inherent in flux calibration are too large to allow a
reliable measurement of the Balmer decrement even if we had been able
to obtain $\Hb$ fluxes; for a Balmer decrement of 10\%, expected for
our mean $E(B-V)=0.10$ mag, we would need 
to measure each line flux with an accuracy of 5\% or less, far better
than our current capabilities.  The issue is further complicated by
the fact that $\Ha$ and $\Hb$ lie in different bands and cannot be
observed simultaneously, so there may be a systematic offset between the flux 
calibrations of the two observations.  It will therefore be difficult to test
the Calzetti model directly.

A comparison of the extinction-corrected SFRs is shown in the right
panel of Figure~\ref{fig:sfrs}.  They are in better agreement than the
uncorrected SFRs,
with $\rm \langle SFR_{\Ha}/SFR_{UV}\rangle = 1.2$ and a reduction in
the scatter of 50\% (1 $\sigma$; again accounting for the lower limits on
four of the $\Ha$ SFRs).  As emphasized above, the extinction
correction is highly sensitive to 
uncertainties in the $G-\cal R$ colors; the errors bars reflect the
errors in $E(B-V)$ determined above, propagated through to the star
formation rates.  Not shown are uncertainties in the extinction law, flux
calibrations, or conversion of flux to star formation rate, all of
which are considerable.  Given these sources of error, and the
uncertainty in the value of $E(B-V)$ that should be used for the
nebular emission, the extinction-corrected SFRs should be taken with
caution. 

\begin{figure*}
\centerline{\epsfxsize=18cm\epsffile{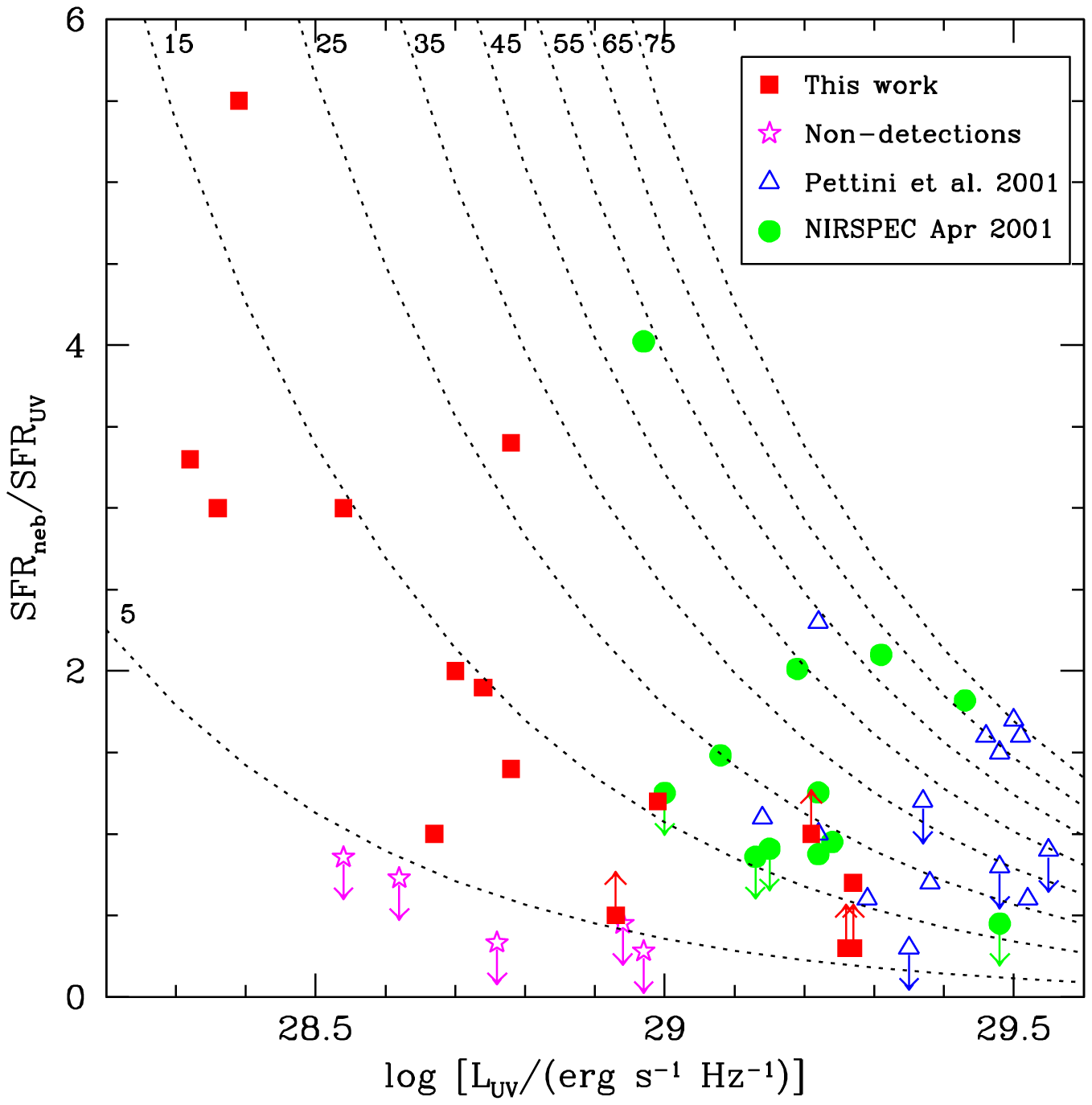}}
\figcaption{A comparison of the ratio of SFRs from nebular line and UV
  continuum emission with UV continuum luminosity.  Red squares are the
  current data set, magenta stars represent upper limits on the ratio of
SFRs for the five objects which we did not detect while successfully
  observing another galaxy on the same slit, blue triangles are
  from the $z\sim3$ sample of \citet{pss+01}, and green circles are
  unpublished $z\sim3$ data from our April 2001 NIRSPEC run.  The
  dotted curves are lines of constant SFR$_{\rm neb}$, and the number
  at the top of each curve is its SFR$_{\rm neb}$ in
  $\msun$~yr$^{-1}$.  UV luminosity is
  computed from the $G$ magnitude for the $z\sim2$ sample; the center
  of the $G$ filter corresponds to $\sim1500$~\AA \space at $z=2.3$.  For the
  $z\sim3$ sample we use the $\cal R$ magnitude, corresponding to
  $\sim1700$~\AA \space at $z=3$.   SFRs for the \citet{pss+01} and
  April 2001 samples were calculated from H$\beta$ emission, 
  assuming $\Ha$/H$\beta=2.75$ and applying the \citet{k98} conversion
  from $\Ha$ to SFR.  Errors are suppressed for clarity, but are
  $\sim$ 25\% in SFR$_{\Ha}$ and 10\% in SFR$_{\mathrm{UV}}$ and $G$
  as discussed in the text.  See \citet{pss+01} for discussion of 
  errors in the SFRs from H$\beta$.
\label{fig:lratio}
}
\end{figure*}

In Figure~\ref{fig:lratio} we plot the ratio $\rm SFR_{neb}/SFR_{UV}$
against the rest frame UV continuum luminosity; none of these quantities
have been corrected for
extinction.  We include data
from \citet{pss+01}, who used H$\beta$ fluxes and the standard ratio
$\Ha/\Hb=2.75$ \citep{o89} to calculate SFRs from recombination lines in
galaxies at $z\sim3$.  We have also included unpublished data from our
NIRSPEC run in April 2001; these are LBGs at $z\sim3$, and star
formation rates have been calculated in the same way as in
\citet{pss+01}.  These data will be discussed in detail in a future
paper.  The dotted curves represent lines of constant nebular line SFR, and the
number at the top of each curve is its SFR$_{\rm neb}$, in
$\msun$~yr$^{-1}$.  We 
see that there is a moderate trend for the UV-faint 
galaxies to have higher nebular line SFRs relative to their UV SFRs,
as might be the case if these objects were more heavily reddened.  
From the curves of constant SFR$_{\rm neb}$ it can be seen that
galaxies with similar nebular line 
luminosities and varying amounts of UV extinction will naturally follow
such a trend.  As noted in \S~\ref{sec:ind}, the UV luminosities of the
galaxies in our sample vary by a factor of 9, while the $\Ha$
luminosities are the same to within a factor of 4; this is consistent
with the idea that the galaxies in our sample have roughly the same
SFR, but differ in the amount of UV extinction.  This model may offer an
explanation for the difference between our results and those of
\citet{pss+01}, who observed no tendency for SFR$_{\rm{H\beta}}$ to be
systematically greater than SFR$_{\rm{UV}}$.  As is apparent from the
figure, the galaxies in their sample are brighter in the UV than all but four
of those presented here, and could plausibly suffer less extinction.
Several caveats are in order, however.  We observe no correlation
between either the $(G-\cal R)$ color or $E(B-V)$ and the ratio of
SFRs; if reddening is indeed the cause of the observed trend, then
UV continuum measurements are not sufficient to quantify it.  We also
note that many objects with faint $\Ha$ emission would fall in
the lower left corner of the plot; this is apparent when we add the
objects we failed to detect to the figure (shown as magenta stars).  We
have plotted only those objects which were placed on the slit with
another galaxy that was detected, so that we know our astrometry was
correct.  We have placed upper limits on their $\Ha$ star formation
rates by assigning a maximum SFR$_{\Ha}$ corresponding to 1 $\sigma$
less than the flux of our weakest detection, and we have calculated UV
SFRs based on their photometry as with the rest of the sample.  It is
clear from this exercise that the absence of data points in the lower
left is a selection effect; such galaxies would have undetectably
small SFRs.  The absence of data points in the upper
right is more significant, as these objects would be easily
detectable; from the curves of constant SFR$_{\rm neb}$, we see that
any galaxies falling here would have extremely large SFRs.  In spite
of these cautions, we believe that this figure is consistent with a
model in which reddening is the primary cause of the discrepancy
between the two SFR indicators.

Changes in the star formation rate on short timescales could also be
reflected in our differing star formation rates, since $\Ha$ emission 
is a more instantaneous measure of the SFR than the UV emission.  The
nebular recombination lines are the reprocessed light of only the most
massive ($M\gtrsim10{\msun}$) and short-lived stars, while the UV
emission probes a wider mass range ($M\gtrsim5{\msun}$).  Therefore a
starburst which has begun in the past $\sim10^8$ yrs will not yet have
reached full UV luminosity and will have an underestimated UV
star formation rate, whereas a decline in star formation will cause an
immediate decrease in $\Ha$ emission as the most massive stars die
off.  In a large sample of galaxies with redshifts $0<z<0.4$,
\citet{ste+00} find that the UV flux indicates a consistently higher SFR than
the $\Ha$, and that the discrepancy is best explained by short bursts
of star formation superimposed on a smooth star formation history.
Such a model could also explain the larger UV SFR of 
a galaxy such as Q1623-BX376; however, this relationship between the UV and
$\Ha$ SFRs is strongest at the fainter end of the \citet{ste+00} sample,
whereas Q1623-BX376 would fall at the bright end.  As noted above,
there were several galaxies which we observed but failed to detect.
This could be explained by a decline in the star formation rate, but
due to the
difficulties presented by the sky background in the IR, the
marginal quality of some of our optical redshifts, and the possibility
of errors in astrometry or the guiding and tracking of the instrument
and telescope,
these objects have not been included in the statistical comparison of SFRs. 

In the following paragraphs we explain why we believe that a young
stellar population is not the primary cause of the discrepancy between
the SFRs.  As we
have no information on the ages of the stellar populations of the
galaxies in our sample, we will assume that they are similar to LBGs
at $z\sim3$, although as we have pointed out above, the samples at $z\sim2$
and $z\sim3$ have different kinematic properties and the $z\sim3$
sample tends to cover brighter UV luminosities.  The stellar
populations of LBGs at $z\sim3$ are now well-studied
\citep{ssa+01,pdf01}, and the \citet{pdf01} sample includes some
galaxies in the range $z=2$--2.5.  Population
synthesis models for a sample of 81 LBGs by \citet{ssa+01} give a
median age since the onset of the most recent episode of star
formation of $t_{\mathrm{sf}}\simeq 320$ Myr, with more than 40\% having
$t_{\mathrm{sf}}>500$ Myr and 25\% having $t_{\mathrm{sf}}<40$ Myr.
We might then expect $\sim25$\% of our sample to have an
underestimated UV SFR; however, the youngest galaxies in the
\citet{ssa+01} sample  are also the most extinguished and have the
highest star formation rates.  Among those with $t_{\mathrm{sf}}<100$
Myr, the mean $E(B-V)$ is 0.27, higher than that of any of the objects
in our sample and  3 $\sigma$ higher than our sample mean of 0.10.  The
mean star formation rate among the same subset is 261 $\msun$~yr$^{-1}$, far
higher than that of any of the objects in our sample even after
correcting for extinction.  Assuming that star-forming galaxies at $z\sim2$ are
similar to those at $z\sim3$, it is therefore unlikely that the
stellar populations of our sample are young enough to account for the
difference in SFRs.

\citet{pdf01} fit a set of detailed models to 33 LBGs
in the Hubble Deep Field North, finding that the age distribution is
strongly dependent on metallicity, IMF, the choice of extinction law,
and the assumed star formation history.  It is possible to vary these
parameters to make the ages young enough to lead to an underestimate
of the UV star formation rate; the youngest ages, $\langle t \rangle
\simeq40$ Myr, are given by a Scalo IMF with 0.2 Z$_{\odot}$.
Although this may be a reasonable estimate for the metallicity of these
objects---\citet{pss+01} find 0.1 to 0.5 Z$_{\odot}$ for galaxies at
$z\sim3$---the theoretical stellar atmospheres used in the population
synthesis models are not well-tested for low metallicities, and the
results should therefore be treated with caution.  More generally,
even ages as young as these cannot fully explain the discrepancy
between the SFRs.  The mean factor of 2.4 difference between the $\Ha$
and UV rates would require the average UV luminosity to have reached
only $\sim40$\% of its full value, which occurs less than 5 Myr
after the beginning of a burst of continuous star formation.  Such an 
extremely young age is unphysical; the time required for a burst 
of star formation to propagate across a galaxy is approximately the 
dynamical timescale, and $t_{dyn} \simeq 30$ Myr for galaxies of the
masses and sizes found in \S~\ref{sec:kin}.  We can state the
timescale argument in another way as well:  the average stellar mass
of our galaxies, $\langle M \rangle \gtrsim 4 \times 10^{10} \msun$,
combined with an assumed age of 2 Gyr gives a characteristic $\dot M
\sim 20 \msun$ yr$^{-1}$, about the same as our mean $\Ha$ SFR of 16 $\msun$
yr$^{-1}$.  This implies that the current SFRs of the galaxies are
similar to their past averages over the last 2 Gyr, and that a current
burst is unlikely.  Assuming an age younger than 2 Gyr, a mass
larger than our lower limit of $4 \times 10^{10} \msun$, or
significant gas recycling results in a current SFR less than the past
average, excluding a current burst even further.

The effects of dust and star formation history are indistinguishable in
individual cases; in the sample taken as a whole, the systematic
depression of SFR$_{\rm UV}$ relative to SFR$_{\rm H\alpha}$ suggests
that extinction is the dominant effect, since variations in star
formation history would induce scatter in the plots rather than
systematic effects.  Our knowledge of star formation and extinction at
high redshift generally supports this conclusion.  A moderate amount
of extinction is indicated by our data, with a mean $E(B-V)$ of 0.10
(corresponding to $A_{1500}\sim$ 1 mag and attenuation by a factor of
$\sim2.5$, using the \citet{cab+00} extinction law); in studies of
LBGs at $z\sim3$, \citet{ssa+01} find a median dust attenuation factor
of $\sim4.5$ at 
$\sim1500$ \AA, while \citet{pdf01} find a factor of 3.0--4.4,
depending on metallicity.  Our results also provide some support for
previous estimates of UV extinction at high redshift: if the $\Ha$
extinction is assumed to be about the same as it is in local galaxies,
a typical factor of 2, and if we assume that the factor of $\sim2.4$
reduction in SFR$_{\rm UV}$ relative to SFR$_{\Ha}$ is due to
extinction, then we obtain a UV extinction factor of $\sim5$, the same
as that applied to the UV luminosity density at $z\sim3$ by
\citet{sag+99}.  We also note that this is in general agreement with
the average UV attenuation factor of 5--6 obtained from studies of the
X-ray luminosity of LBGs at $z\sim3$ \citep{nma+02}. In summary, while
we cannot rule out the effects of star formation history entirely, our
results are consistent 
with other estimates of extinction in galaxies at high redshift, and
such extinction naturally explains the differences we see in the $\Ha$-
and UV-derived star formation rates.

\section{Summary and Conclusions} 
\label{sec:disc}

We have presented $\Ha$ spectroscopy of 16 galaxies in the redshift
range $2.0 < z < 2.6$; this is so far the largest sample
of near-IR spectra of galaxies at these redshifts.  The galaxies were
selected based on their broadband rest-frame UV colors, using an
adaptation of the technique used to select Lyman break galaxies at
$z\sim3$.  Those observed here are drawn from a large sample of such
galaxies, with redshifts already confirmed; because proximity to a
QSO sight-line was the primary selection criterion for near-IR observation, we
believe the 16 galaxies presented here to be representative of the
sample as a whole.  We have analyzed the spectra in order to determine
the kinematic and star-forming properties of the galaxies, and we
reach the following conclusions.

1.  Six of the 16 galaxies show spatially extended, tilted $\Ha$
    emission lines, such as would be produced by ordered rotation.
    Rotation curves for these galaxies show a mean velocity of
    $\sim150$ \kms at a mean radius of $\sim6$ kpc; these are lower
    limits obtained by taking half of the total range in both velocity
    and distance.  Measuring from the spatial location of the
    continuum and the dynamical center of the lines, we obtain a
    maximum velocity of $\sim240$ \kms and a maximum radius of 10 kpc
    in the most extreme cases.  We have obtained archival \textit{HST}
    images for two of these galaxies, and they appear to be
    morphologically irregular, as do all of the other galaxies in our
    sample for which we have such images.  Because of their chaotic
    morphologies, we have not attempted to model any corrections to
    the rotation curves.  We have used the lower limits on the
    rotational velocity and radius of each galaxy to derive a
    dynamical mass; we obtain a mean of $\langle M \rangle \geq 4 \times
    10^{10} \msun$.  Because $\Ha$ emission probes only the central
    star-forming regions of the galaxies, we expect their total halo
    masses to be several times larger.  These results are in general
    agreement with the predictions of models of hierarchical galaxy
    formation for LBGs at $z\sim3$.

2.  Values of the one-dimensional velocity dispersion $\sigma$ range
    from 50 to 260 \kms, with a mean of $\sim110$ \kms.
    Assuming that the line widths are due to gravitational motions in
    the potentials of the galaxies, the mean virial mass implied is $2
    \times 10^{10} \msun$; this is in general agreement with the masses
    we obtain from the rotation curves.  We consider other possible
    origins for the broadening of the lines, including large-scale
    outflows, mergers and interactions.

3.  Both the rotational velocity $v_c$ and the velocity dispersion
    $\sigma$ tend to be larger at $z\sim2$ than at $z\sim3$.  We see
    evidence of rotation in $\sim40$\% of our sample, whereas
    \citet{pss+01} found such evidence in only $\sim10$\% of a sample
    of similar size at $z\sim3$.
    Furthermore, we find rotational velocities of $\sim150$ \kms, as
    compared to $\sim50$ \kms at $z\sim3$.  Our mean value of
    $\sigma$, $\sim110$ \kms, is
    $\sim60$\% larger than the value found at $z\sim3$ by \citet{pss+01}.
    We have considered possible selection effects that may explain
    these systematic differences, but have not found a convincing
    explanation.  It may be that the redshift dependence of surface
    brightness allows us to sample to larger radii at $z\sim2$, or
    that our photometric selection criteria pick out different
    populations of galaxies at $z\sim2$ and $z\sim3$.  It is also
    possible that the effect is real and reflects the growth of disks
    between these two epochs. 

4.  We use the $\Ha$ luminosity to calculate the star formation rates
    of the galaxies, and compare these to the SFRs derived from the
    rest-frame UV continuum luminosity.  We use the calibrations of
    \citet{k98} in both cases.  We obtain a mean $\rm SFR_{\Ha}$ of 16
    $\msun$~yr$^{-1}$, and a mean $\rm SFR_{\Ha}/SFR_{UV}$ ratio of
    2.4.  After correcting both luminosities for extinction using the
    \citet{cab+00} extinction law, we find $\rm
    SFR_{\Ha}/SFR_{UV}=1.2$, with a 50\% reduction in scatter.  We
    discuss the effects of extinction and star formation history on
    the SFRs, and conclude that extinction is the more likely
    explanation for their discrepancy.  We also see a moderate
    correlation between the ratio $\rm SFR_{\Ha}/SFR_{UV}$ and the UV
    luminosities of the galaxies, such that UV-faint galaxies have a
    higher $\rm SFR_{\Ha}/SFR_{UV}$ ratio.  Such an effect could be
    produced if the fainter galaxies undergo more extinction.

5.  Finally, we expect that many of the points discussed here will
    become clearer as the sample of near-IR observations of galaxies
    at these redshifts grows.  The photometric technique for selecting
    galaxies at $z\sim2$ has so far produced hundreds of galaxies with
    confirmed redshifts in this range, and further observations of
    their kinematics, line fluxes, and morphologies will shed light on
    star formation, extinction, and the formation of disks at high
    redshift. 

\acknowledgements

We would like to thank the referee for many helpful
comments and suggestions.  We also thank the staffs at
the Keck and VLT observatories for their competent assistance with the
observations.  CCS, DKE, AES and MPH have been supported by grant
AST-0070773 from the U.S. National Science Foundation and by the David
and Lucile Packard Foundation. KLA acknowledges support from the
Harvard Society of Fellows.  Finally, we wish to extend special thanks
to those of Hawaiian ancestry on whose sacred mountain we are privileged
to be guests.  Without their generous hospitality, most of the
observations presented herein would not have been possible.



\begin{deluxetable}{l l l l l l l l l l}
\tablewidth{0pt}
\tablecaption{Galaxies Observed\label{tab:obs}}
\rotate
\tablehead{
\colhead{Galaxy} & 
\colhead{R.A. (J2000)} & 
\colhead{Dec. (J2000)} & 
\colhead{$z_{\rm Ly\alpha}$\tablenotemark{a}} & 
\colhead{$z_{\rm abs}$\tablenotemark{b}} & 
\colhead{$z_{\Ha}$\tablenotemark{c}} & 
\colhead{$\cal R$} & 
\colhead{$G-\cal R$} & 
\colhead{Exposure (s)} & 
\colhead{Telescope/Instrument} 
}
\startdata
CDFb-BN88 & 00:53:52.87 & 12:23:51.25 & --- & 2.263 &  2.2615 & 23.14 & 0.29 & 12$\times$720 & VLT 1/ISAAC\\
Q0201-B13   & 02:03:49.25 & 11:36:10.58 & --- & 2.167 & 2.1663  & 23.34   & 0.02    & 16$\times$720  & VLT 1/ISAAC\\
Westphal-BX600\tablenotemark{d}  & 14:17:15.55 & 52:36:15.64 & --- & --- & 2.1607  & 23.94   & 0.10    & 5$\times$900   & Keck II/NIRSPEC\\
Q1623-BX376 & 16:25:05.63  & 26:46:49.12 & 2.415 & 2.408 & 2.4085  & 23.31   & 0.24    & 4$\times$900   & Keck II/NIRSPEC\\
Q1623-BX428 & 16:25:48.42 & 26:47:40.24 & --- & 2.053 & 2.0538\tablenotemark{e}  & 23.95   & 0.13    & 4$\times$900   & Keck II/NIRSPEC\\
Q1623-BX432 & 16:25:48.74 & 26:46:47.05 & 2.187 & 2.180 & 2.1817  & 24.58   & 0.10    & 4$\times$900   & Keck II/NIRSPEC\\
Q1623-BX447 & 16:25:50.38 & 26:47:14.07 & --- & 2.149 & 2.1481  & 24.48   & 0.17    & 4$\times$900   & Keck II/NIRSPEC\\
Q1623-BX449 & 16:25:50.55 & 26:46:59.63 & --- & 2.417 & 2.4188  & 24.86   & 0.20    & 4$\times$900   & Keck II/NIRSPEC\\
Q1623-BX511 & 16:25:56.10 & 26:44:44.38 & --- & 2.246 & 2.2421\tablenotemark{e}  & 25.37   & 0.42    & 4$\times$900   & Keck II/NIRSPEC\\
Q1623-BX522 & 16:25:55.76 & 26:44:53.17 & --- & 2.476 & 2.4757  & 24.50   & 0.31    & 4$\times$900   & Keck II/NIRSPEC\\
Q1623-MD107 & 16:25:53.88 & 26:45:15.19 & 2.543 & 2.536 & 2.5373  & 25.35   & 0.12    & 4$\times$900   & Keck II/NIRSPEC\\
Q1700-BX691 & 17:01:05.99 & 64:12:10.27 & --- & 2.189 & 2.1895  & 25.33   & 0.22    & 4$\times$900   & Keck II/NIRSPEC\\
Q1700-BX717 & 17:00:57.00  & 64:12:23.71 & 2.438 & --- & 2.4353  & 24.78   & 0.20    & 4$\times$900   & Keck II/NIRSPEC\\
Q1700-MD103 & 17:01:00.20 & 64:11:56.00 & --- & 2.308 & 2.3148  & 24.23   & 0.46    & 900+600 & Keck II/NIRSPEC\\
Q1700-MD109 & 17:01:04.48 & 64:12:09.28  & 2.295 & 2.297 & 2.2942  & 25.46   & 0.26    & 4$\times$900   & Keck II/NIRSPEC\\
SSA22a-MD41 & 22:17:39.97 & 00:17:11.04 & --- & 2.173 & 2.1713  & 23.31   & 0.19    & 15$\times$720  & VLT 1/ISAAC
\enddata
\tablenotetext{a}{Vacuum heliocentric redshift of Ly$\alpha$ emission
  line, when present.}
\tablenotetext{b}{Vacuum heliocentric redshift from rest-frame UV
  interstellar absorption lines.}
\tablenotetext{c}{Vacuum heliocentric redshift of H$\alpha$ emission
  line.}
\tablenotetext{d}{We have not yet obtained a rest-frame UV spectrum of Westphal-BX600.}
\tablenotetext{e}{The $\Ha$ redshifts of the galaxies Q1623-BX428 and
  Q1623-BX511 are somewhat uncertain due to the presence of strong sky
  lines near $\Ha$.}
\end{deluxetable}

\begin{deluxetable}{l l l c l l}
\tablecaption{Kinematics\label{tab:kin}}
\tablewidth{0pt}
\tablehead{\colhead{Galaxy} & \colhead{$z_{{\rm H}\alpha}$\tablenotemark{a}} &
\colhead{$\sigma$} & \colhead{$v_c$\tablenotemark{b}} & \colhead{M$_{vir}$\tablenotemark{c}} &
\colhead{M$_{dyn}$\tablenotemark{d}}\\
\colhead{ } & \colhead{ } & \colhead{(km s$^{-1}$)} & \colhead{(km
  s$^{-1}$)} & \colhead{($\msun$)} & \colhead{($\msun$)}}
\startdata
CDFb-BN88 & 2.2615 & $96\pm46$ & --- & $1.7 \times 10^{10}$ & ---\\
Q0201-B13 & 2.1663 & $62\pm29$ & --- & $7.2 \times 10^{9}$ & ---\\
Westphal-BX600 & 2.1607 & $181\pm24$ & $\sim210$ & $6.2 \times 10^{10}$ & $6.0 \times 10^{10}$\\
Q1623-BX376a & 2.4085 & $261\pm72$ & --- & $1.3 \times 10^{11}$ & ---\\
Q1623-BX376b & 2.4085 & $<224$ & --- & $<9.4 \times 10^{10}$ & ---\\
Q1623-BX428\tablenotemark{e} & 2.0538 & --- & --- & --- & ---\\
Q1623-BX432 & 2.1817 & $51\pm22$ & --- & $5.0 \times 10^9$ & ---\\
Q1623-BX447 & 2.1481 & $174\pm18$ & $\sim160$ & $5.8 \times 10^{10}$ & $3.0 \times 10^{10}$\\
Q1623-BX449 & 2.4188 & $141\pm94$ & --- & $3.7 \times 10^{10}$ & ---\\
Q1623-BX511 & 2.2421 & $152\pm47$ & $\sim80$ & $4.4 \times 10^{10}$ & $4.6 \times 10^9$\\
Q1623-BX522 & 2.4757 & $<44$ & --- & $<3.8 \times 10^{10}$ & ---\\
Q1623-MD107 & 2.5373 & $<42$ & --- & $<2.9 \times 10^{10}$ & ---\\
Q1700-BX691 & 2.1895 & $170\pm18$ & $\sim220$ &$5.5 \times 10^{10}$ & $7.0 \times 10^{10}$\\
Q1700-BX717 & 2.4353 & $<60$ & --- & $<1.3 \times 10^{10}$ & ---\\
Q1700-MD103 & 2.3148 & $75\pm21$ & $\sim100$ & $1.1 \times 10^{10}$ & $1.6 \times 10^{10}$\\
Q1700-MD109 & 2.2942 & $87\pm35$ & --- & $1.5 \times 10^{10}$ & ---\\
SSA22a-MD41 & 2.1713 & $107\pm15$ & $\sim150$ & $2.2 \times 10^{10}$ & $4.2 \times 10^{10}$
\enddata
\tablenotetext{a}{Vacuum heliocentric redshift of H$\alpha$ emission line.}
\tablenotetext{b}{Minimum rotational velocity, $(v_{max}-v_{min})/2$.}
\tablenotetext{c}{Masses calculated from the velocity dispersion.}
\tablenotetext{d}{Minimum masses derived from rotational velocities when
  available.}
\tablenotetext{e}{Sky line contamination prevented a measurement of $\sigma$.}
\end{deluxetable}

\begin{deluxetable}{l r r c r r c r r r r r}
\tablecaption{Fluxes and Star Formation Rates\label{tab:sfr}}
\tablehead{\colhead{Galaxy} & \colhead{$z_{{\rm
	H}\alpha}$\tablenotemark{a}} & \colhead{$\cal R$} &
  \colhead{$G-{\cal R}$} & \colhead{$F_{{\rm
	H}\alpha}$\tablenotemark{b}} & \colhead{$L_{{\rm
	H}\alpha}$\tablenotemark{c}} &
  \colhead{$E(B-V)$\tablenotemark{d}} & \colhead{SFR$_{{\rm
	H}\alpha}$\tablenotemark{e}} & \colhead{SFR$_{{\rm
	H}\alpha}$\tablenotemark{f}} & \colhead{SFR$_{\rm
      UV}$\tablenotemark{g}} & \colhead{SFR$_{\rm
      UV}$\tablenotemark{h}} & \colhead{$\frac{\rm{SFR}_{{\rm
	  H}\alpha}}{\rm{SFR}_{UV}}$\tablenotemark{i}} 
}
\startdata
CDFb-BN88   & 2.2615  & 23.14 & 0.29 & 2.6 & 1.0 & 0.146 & $>8$ & $>14$  & $26\pm3$ & $106^{+31}_{-24}$  & $>0.3$\\
Q0201-B13   & 2.1663  & 23.34 & 0.02 & 2.4 & 0.8 & 0.004 & $>7$ & $>7$  & $26\pm3$ & $27^{+6}_{-5}$  & $>0.3$\\
Westphal-BX600 & 2.1607  & 23.94 & 0.10 & 6.3 & 2.2 & 0.048 & $17\pm4$ & $21^{+3}_{-3}$ & $14\pm1$ & $22^{+11}_{-7}$  & 1.2\\
Q1623-BX376 & 2.4085 & 23.31 & 0.24 &  5.3 & 2.4 & 0.111 & $19\pm5$ & $29^{+3}_{-3}$ & $26\pm3$ & $84^{+27}_{-20}$ & 0.7\\
Q1623-BX428 & 2.0538  & 23.95 & 0.13 & 2.7 & 0.8 & 0.073 & $>6$ & $>9$  & $12\pm1$ & $25^{+12}_{-8}$  & $>0.5$\\
Q1623-BX432 & 2.1817  & 24.58 & 0.10 & 5.4 & 1.9 & 0.048 & $15\pm4$ & $18^{+4}_{-3}$  &  $8\pm1$ & $13^{+8}_{-5}$  & 1.9\\
Q1623-BX447 & 2.1481  & 24.48 & 0.17 & 5.6 & 1.9 & 0.082 & $15\pm4$ & $21 ^{+5}_{-4}$ &  $8\pm1$ & $18^{+12}_{-7}$  & 1.9\\
Q1623-BX449 & 2.4188  & 24.86 & 0.20 & 1.8 & 0.8 & 0.094 & $6\pm2$ & $9^{+2}_{-2}$  &  $6\pm1$ & $18^{+15}_{-8}$  & 1.0\\
Q1623-BX511 & 2.2421  & 25.37 & 0.42 & 3.4 & 1.3 & 0.194 & $10\pm3$ & $22^{+8}_{-6}$  &  $3\pm0.3$ & $22^{+28}_{-12}$  & 3.3\\
Q1623-BX522 & 2.4757  & 24.50 & 0.31 & 2.8 & 1.3 & 0.132 & $11\pm3$ & $18^{+4}_{-3}$  &  $8\pm1$ & $35^{+26}_{-15}$  & 1.4\\
Q1623-MD107 & 2.5373  & 25.35 & 0.12 & 3.7 & 1.9 & 0.043 & $15\pm4$ & $18^{+6}_{-4}$  &  $5\pm1$ &  $8^{+9}_{-4}$  & 3.0\\
Q1700-BX691 & 2.1895  & 25.33 & 0.22 & 7.7 & 2.8 & 0.108 & $22\pm6$ & $33^{+12}_{-9}$  &  $4\pm0.4$ & $10^{+12}_{-7}$  & 5.5\\
Q1700-BX717 & 2.4353  & 24.78 & 0.20 & 3.8 & 1.8 & 0.087 & $14\pm4$ & $20^{+5}_{-4}$  &  $7\pm1$ & $18^{+15}_{-8}$  & 2.0\\
Q1700-MD103 & 2.3148  & 24.23 & 0.46 & 8.2 & 3.4 & 0.224 & $27\pm7$ & $64^{+13}_{-11}$  &  $8\pm1$ & $88^{+56}_{-35}$  & 3.4\\
Q1700-MD109 & 2.2942  & 25.46 & 0.26 & 2.8 & 1.1 & 0.124 & $9\pm2$ & $14^{+5}_{-4}$  &  $3\pm0.3$ & $12^{+14}_{-6}$  & 3.0\\
SSA22a-MD41 & 2.1713  & 23.31 & 0.19 & 7.9 & 2.8 & 0.097 & $>22$ & $>32$  & $23\pm2$ & $61^{+20}_{-15}$  & $>1.0$\\
\tableline
Mean value\tablenotemark{j} & 2.2787 & 24.37 & 0.21 & 4.6 & 1.8 & 0.101 & 16 & 26 & 12 & 35 & 2.4
\enddata
\tablenotetext{a}{Vacuum heliocentric redshift of H$\alpha$ line.}
\tablenotetext{b}{Line flux in units of $10^{-17}\mbox{ erg s}^{-1}\mbox{ cm}^{-2}$.}
\tablenotetext{c}{Luminosity in units of $10^{42}\mbox{ erg s}^{-1}$.}
\tablenotetext{d}{From $G-\cal R$ colors, corrected as described in \S~\ref{sec:sfrs}.}
\tablenotetext{e}{SFR in M$_{\odot}\mbox{ yr}^{-1}$ from H$\alpha$ luminosity, uncorrected 
for extinction.}
\tablenotetext{f}{SFR in M$_{\odot}\mbox{ yr}^{-1}$ from H$\alpha$ luminosity, corrected 
for extinction.}
\tablenotetext{g}{SFR in M$_{\odot}\mbox{ yr}^{-1}$ from $G$ magnitude, uncorrected for extinction.}
\tablenotetext{h}{SFR in M$_{\odot}\mbox{ yr}^{-1}$ from $G$ magnitude, corrected for extinction.}
\tablenotetext{i}{Ratio of uncorrected SFRs.}
\tablenotetext{j}{For those quantities containing lower limits,
  statistics are computed using survival analysis as discussed in \S~\ref{sec:sfrs}.} 
\tablecomments{$H_0=70\;{\rm km}\;{\rm s}^{-1}\;{\rm Mpc}^{-1}$, $\Omega_m=0.3$, and $\Omega_{\Lambda}=0.7$.}
\end{deluxetable}

\end{document}